\documentclass[aps, prl, reprint, superscriptaddress, longbibliography, amsmath, amssymb, amsfonts, nobibnotes, nofootinbib, floatfix]{revtex4-2}

\usepackage[utf8]{inputenc}
\usepackage[english]{babel}
\usepackage{graphicx}
\usepackage{hyperref}

\usepackage{bm}
\usepackage{mathbbol}
\usepackage{amsthm}
\usepackage{mathrsfs}
\usepackage{commath} 
\usepackage[mathcal]{euscript}
\usepackage{latexsym}
\usepackage{float}
\usepackage{subfig}
\usepackage{subcaption}

\graphicspath{{images/}}

\newcommand{\be}{\begin{equation}}
\newcommand{\ee}{\end{equation}}
\newcommand{\eps}{\varepsilon}
\newcommand{\om}{\omega}

\newcommand{\Gloc}{G_\text{loc}}
\newcommand{\Gglob}{G_\text{glob}}

\usepackage{mathtools}
\DeclarePairedDelimiter\corrfunc{\langle}{\rangle}
\DeclarePairedDelimiter\bra{\langle}{\rvert}
\DeclarePairedDelimiter\ket{\lvert}{\rangle}

\begin{document}

\title{Random matrix theory signatures in free field theory}

\author{Dmitry S. Ageev}
\email{ageev@mi-ras.ru}
\affiliation{Steklov Mathematical Institute of Russian Academy of Sciences, 8 Gubkina St., Moscow 119333, Russia}
\affiliation{Institute for Theoretical and Mathematical Physics, Lomonosov Moscow State University, 119991 Moscow, Russia}

\author{Vasilii V. Pushkarev}
\email{pushkarev@mi-ras.ru}
\affiliation{Steklov Mathematical Institute of Russian Academy of Sciences, 8 Gubkina St., Moscow 119333, Russia}

\date{\today}

\begin{abstract}
    We show that, within a finite window of parameter space, random matrix theory (RMT) statistics emerge in observables of a finite-volume massive free scalar field theory after a local operator quench. The spacing-ratio distribution of two-point-function extremum locations is close to the Gaussian orthogonal ensemble statistics. An extrema-based form factor exhibits a dip--ramp--plateau structure characteristic of RMT. By contrast, the standard spectral form factor shows no ramp, consistent with the underlying free spectrum, while a global quench yields qualitatively different statistics.
\end{abstract}

\maketitle

\textit{Introduction.---}Defining chaos and sharpening our understanding of it are long-standing goals in classical and quantum mechanics. Beyond its role in microscopic physics, quantum chaos is tightly connected to problems in black-hole physics, string theory, condensed matter, and quantum complexity. Important relations between chaos, gravity, conformal field theory, and the holographic correspondence were established in seminal papers~\cite{Sekino:2008he, Shenker:2013pqa, Maldacena:2015waa} and have continued to be explored further in many subsequent studies, for example, \cite{Brandino:2010sv, Roberts:2014ifa, Dyer:2016pou, Cotler:2016fpe, Grozdanov:2017ajz, Gharibyan:2019sag, Cubrovic:2019qee, Stanford:2019vob, Rosenhaus:2020tmv, Ageev:2020qox, Srdinsek:2020bpq, Gross:2021gsj, Ageev:2021poy, Delacretaz:2022ojg, Bianchi:2022mhs, Das:2022evy, Negro:2022hno, Savic:2024ock, Sonnenschein:2025jlc}.

In many quantum systems believed to exhibit chaotic dynamics, spectral fluctuations display universal statistical properties described by random matrix theory (RMT)~\cite{Bohigas84, Kos:2017zjh}. An important question is how far this universality extends beyond the spectrum itself and whether it can be directly detected in observables~\cite{Gharibyan:2019sag, Bianchi:2022mhs, Bianchi:2023uby, Bianchi:2024fsi, Bianchi:2025kna}.

We show that RMT-like statistics appear in observables in an unexpectedly simple setting: a local operator quench in a finite-volume massive free scalar field theory. This is conceptually surprising because the post-quench dynamics is governed by a free Hamiltonian, so conventional spectral diagnostics are not expected to display RMT universality.

An instantaneous quantum quench provides a minimal probe of nonequilibrium dynamics: a system is prepared in an initial state and then evolved unitarily under a Hamiltonian for which that state is not an eigenstate. A local quench generates a spatially localized distribution of correlations near the quench point~\cite{Calabrese:2007mtj}. Here we consider the local operator quench~\cite{Nozaki:2014hna}, in which a nonequilibrium state is prepared by acting with a local operator on the vacuum. Most studies of local operator quenches in quantum field theory have focused on conformal field theories and holography~\cite{Nozaki:2013wia, Nozaki:2014hna, He:2014mwa, Caputa:2014vaa, Asplund:2014coa, Caputa:2014eta, Calabrese:2016xau, Ageev:2018nye}; the local operator quench in massive scalar field theories was studied in~\cite{Ageev:2022kpm}.

Specifically, for the two-point correlation function following a local operator quench in two-dimensional massive free scalar field theory on a finite spatial circle, we find that:
\begin{enumerate}
    \item the spacing-ratio statistics of extrema in the correlation function are close to those of Gaussian ensembles of random matrices;
    \item the spacetime correlation form factor, constructed analogously to the spectral form factor but based on extrema statistics, shows a dip--ramp--plateau structure.
\end{enumerate}
These signatures arise in an intermediate window of model parameter space and cross over to more regular behavior outside it. By contrast, the conventional spectral form factor defined from the partition function does not exhibit a ramp, as expected for free theories.

Finally, we compare these features with those of the boundary global quench~\cite{Calabrese:2006rx, Calabrese:2007rg}. Unlike a local quench, a global quench produces a spatially homogeneous excitation. This nonequilibrium protocol is widely used, for example, in holography~\cite{Danielsson:1999fa, Balasubramanian:2010ce, Balasubramanian:2011ur, Ageev:2017wet}.  We find that the statistics of observables differ significantly from those of the local-quench setup.

\textit{Setup.---}We study a two-dimensional massive free scalar field theory in a finite volume with the Euclidean action
\be
    S = \frac{A}{2}\int\,d\tau\,\int_{-L/2}^{L/2} dx \left((\partial_\tau\phi)^2 + (\partial_x\phi)^2 + m^2\phi^2\right),
    \label{eq:model}
\ee
where $\tau$ is Euclidean time and $m$ is the field mass. In the numerical results below, we set the normalization constant to $A = 1/(4\pi)$. The underlying geometry is a cylinder of circumference~$L$, which imposes periodic boundary conditions, $\phi(\tau, x) = \phi(\tau, x + L)$. This implies that the momentum modes take discrete values, ${k_n = 2\pi n/L}$, $n \in \mathbb{Z}$, and that the two-point correlation function is also periodic, ${G(\tau, x) = G(\tau, x + L)}$. Explicitly, it is given by
\be
    \begin{aligned}
        G(\tau, x) = \frac{1}{AL}\sum_n\frac{e^{-\om_n\sqrt{\tau^2} + ik_n x}}{2\om_n},
    \end{aligned}
    \label{eq:equilib_2point}
\ee
where $\om_n$ is given by the finite-volume dispersion relation, $\om_n = \sqrt{k_n^2 + m^2}$. The Lorentzian correlator can be obtained by analytic continuation ${\tau \to it}$; note that $\sqrt{\tau^2}$ encountered in~\eqref{eq:equilib_2point} determines how to perform this continuation when $\tau$ has both non-zero real and imaginary parts.\footnote{This is what occurs in the local operator quench setup; for details, refer to Appendices~A and~B of~\cite{Ageev:2022kpm}.} After analytic continuation, the exponentials in the series become oscillatory and the low-lying modes dominate. We evaluate the series numerically, truncating it at ${n_\text{max} = 250}$.

\textit{Local operator quench.} Insertion of a local operator~$O$ prepares a locally excited state~\cite{Nozaki:2014hna},
\be
    \ket{\Psi(t)} = \mathcal{N}\,e^{-iH(t - t_0)} e^{-\eps H} O(t_0, x_0)\ket{0},
    \label{eq:excited_state}
\ee
where $H$~is the Hamiltonian. The parameter~$\eps$ regulates ultraviolet modes, while the normalization factor $\mathcal{N}$ ensures that the state has unit norm. In what follows, we take the field operator $\phi$ as the simplest example of $O$. The real-time dynamics of the equal-time two-point correlation function after the quench is given by
\be
    \begin{aligned}
        \Gloc(t, x) &\equiv \frac{\bra{0}\phi(i\eps, 0)\phi(t, x)\phi(t, 0)\phi(-i\eps, 0)\ket{0}}{\bra{0}\phi(i\eps, 0)\phi(-i\eps, 0)\ket{0}} - \\
        & -\bra{0}\phi(t, x)\phi(t, 0)\ket{0},
    \end{aligned}
    \label{eq:Gloc}
\ee
where the equilibrium background has been subtracted.

\begin{figure}[t]
    \includegraphics[width=1.\columnwidth]{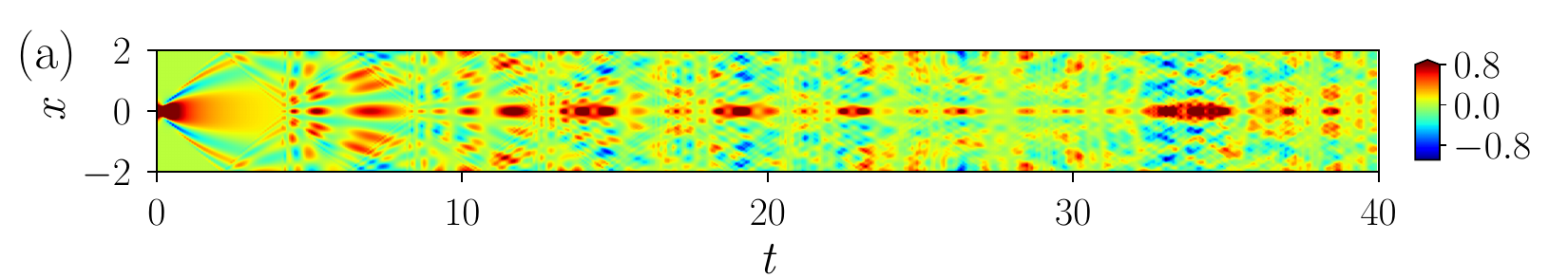}
    \includegraphics[width=1.\columnwidth]{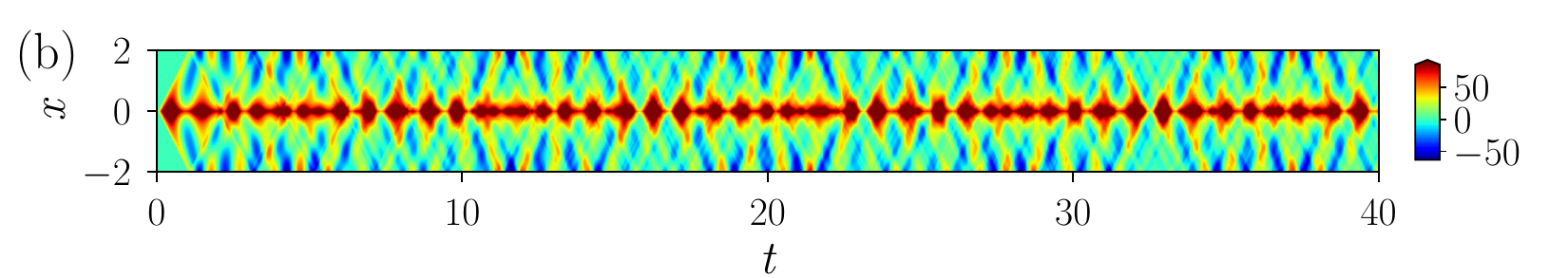}
    \caption{\label{fig:G_xt}Two-point correlation function after: \textit{(a)}~local operator quench, ${L = 4}$, ${m = 10}$, ${\eps = 0.05}$; \textit{(b)}~boundary global quench, ${L = 4}$, ${m = 3}$, ${\tau_0 = 0.01}$.}
\end{figure}

\begin{figure}[t]
    \includegraphics[width=0.45\columnwidth]{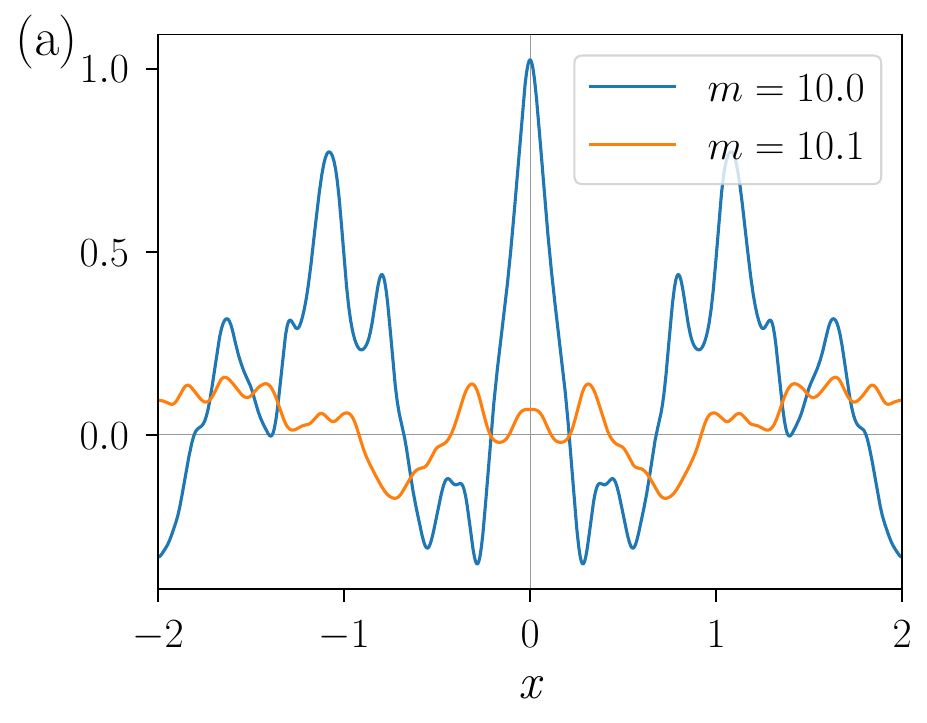}
    \includegraphics[width=0.45\columnwidth]{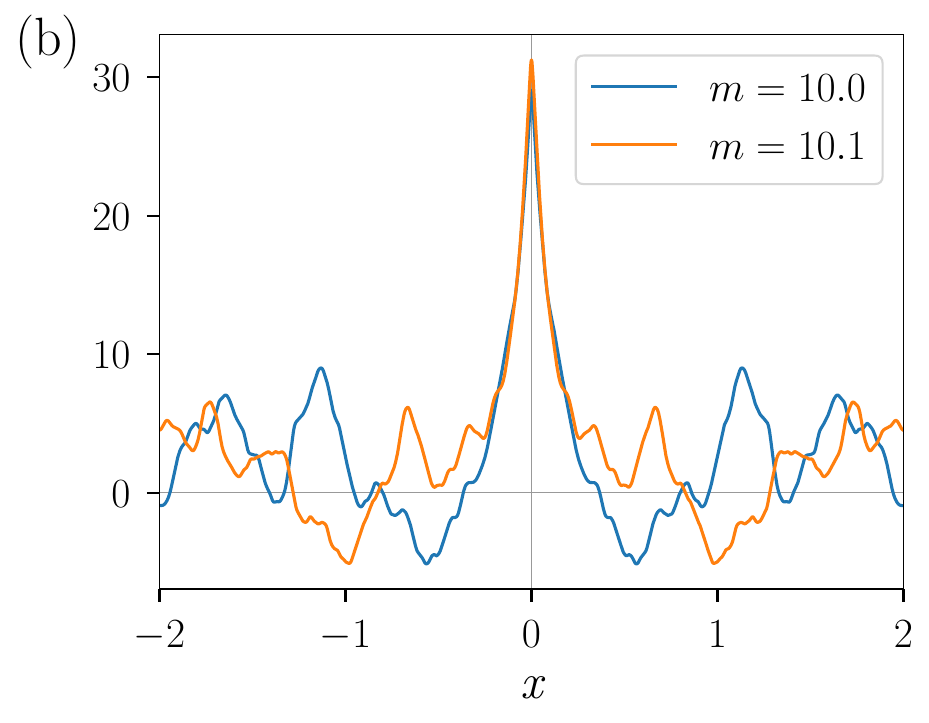}
    \caption{\label{fig:profiles_m}Two-point correlation function for slightly different masses~$m$ at a fixed large time moment (${t = 500}$) after: \textit{(a)} local operator quench; \textit{(b)} boundary global quench. The parameters are: ${L = 4}$, ${m = 10}$ (blue line) and ${m = 10.1}$ (orange line), ${\eps = 0.05}$ (for local quench) and ${\tau_0 = 0.01}$ (for global quench).}
\end{figure}

As shown in Fig.~\ref{fig:G_xt}(a), an irregular structure emerges in the dynamics of~\eqref{eq:Gloc}. For a fixed value of the regulator $\eps$, this structure depends on both the scalar field mass and the circumference of the cylinder; in particular, larger mass enhances the erratic patterns. Comparing the pairs of sample spatial profiles at a fixed late time in Fig.~\ref{fig:profiles_m}, we see that changes in the mass of the relative order $10^{-2}$ can significantly influence the dynamics when the mass is large (and similarly for changes in the circumference, see Fig.~\ref{fig:profiles_L} in the Supplemental Material).

\textit{Boundary global quench.} A global nonequilibrium excitation is prepared by restricting the path-integral geometry of the field theory to a slab of width~$\tau_0$ in Euclidean time, with boundary conditions imposed at the slab boundaries $\pm\tau_0/2$~\cite{Calabrese:2006rx, Sotiriadis:2010si}. For the finite-volume case with Dirichlet slab boundary conditions, the real-time nonequilibrium two-point function is given by~\cite{Calabrese:2007rg, Rajabpour:2014osa}
\be
    \begin{aligned}
        \Gglob(t, x) &\equiv \corrfunc{\phi(t, x)\phi(t, 0)}_\text{slab} - \corrfunc{\phi(0, x)\phi(0, 0)}_\text{slab} = \\
        & = \frac{1}{AL}\sum_n e^{ik_n x}\frac{\sin^2(\om_n t)}{\om_n\sinh(\om_n\tau_0)},
    \end{aligned}
    \label{eq:Gglob}
\ee
where $k_n$ and $\om_n$ are the discrete modes defined as in \eqref{eq:equilib_2point}. The initial background is subtracted.

Sample spatiotemporal profiles are shown in Figs.~\ref{fig:G_xt}(b) and \ref{fig:profiles_m}(b), in comparison with the local-quench case. At first sight, both cases appear erratic; the statistical analysis of observables, however, reveals a marked difference associated with the breaking of spatial translational invariance by the local operator insertion.

\textit{I: Spacing-ratio statistics of extrema of the correlation function.---}It is conjectured that the spectral-fluctuation statistics of a quantum-chaotic system are closely related to those of the Gaussian random matrix ensembles~\cite{Bohigas84}. Such statistics include the nearest-neighbor level-spacing distribution, which measures correlations between adjacent levels in the spectrum. The appropriate ensemble is determined primarily by antiunitary symmetries, for example, the Gaussian orthogonal ensemble (GOE) when time-reversal symmetry is present and the Gaussian unitary ensemble (GUE) when it is broken~\cite{Haake:2010fgh}. By contrast, the level-spacing distribution for integrable systems closely follows the Poisson distribution~\cite{BerryTabor77}.

More robust than the spacings $\delta_i = \lambda_{i + 1} - \lambda_i$ for a spectrum $\{\lambda_i\}$ are their ratios~\cite{Oganesyan:2007wpd, Atas:2013gvn}, $r_i = \delta_{i}/\delta_{i - 1}$. Ratios do not require unfolding, namely the subtraction of the smooth part of the level density, a procedure that must be tuned for each system. The analytical expressions for the spacing-ratio probability density functions (PDFs) for GOE ($\alpha = 1$) and GUE ($\alpha = 2$) are~\cite{Atas:2013gvn}
\be
    f_\alpha(r) = \frac{1}{Z_\alpha}\frac{(r + r^2)^\alpha}{(1 + r + r^2)^{(1 + 3\alpha/2)}},
    \label{eq:pdfs}
\ee
where $Z_1 = 8/27$ and $Z_2 = 4\pi/(81\sqrt{3})$.

In quantum mechanics, the level-spacing distribution is straightforward to calculate once the Hamiltonian eigenvalues are known. In quantum field theory, however, this task is much more difficult because the spectrum is often not explicitly accessible. One possible route is to approximate the spectrum using Hamiltonian truncation~\cite{Srdinsek:2020bpq, Sonnenschein:2025jlc}. A different approach was developed in~\cite{Bianchi:2022mhs, Bianchi:2023uby} for scattering amplitudes and applied in particular to the scattering of highly excited string states. It analyzes the successive ratios of spacings between amplitude peaks as an alternative chaotic diagnostic. This observable is closely related to spacing-ratio statistics for the nontrivial zeros of the Riemann zeta function on the critical line, which are numerically observed to agree with GUE statistics~\cite{berry2005riemann, odlyzko1987distribution, Atas:2013gvn}.

Here, we use peak-statistics analysis for the nonequilibrium two-point functions \eqref{eq:Gloc} and \eqref{eq:Gglob}. For a set of slices at fixed spatial coordinate, we locate the extrema of the correlator as a function of time over the interval from the quench time up to a cutoff time. We then compute spacing ratios for each slice and combine them into a single sample across all slices. From this sample, we construct the spacing-ratio distribution normalized to unity, which we refer to as the collective statistics of extrema over all the chosen slices.

\begin{figure}[t]
    \includegraphics[width=0.49\columnwidth]{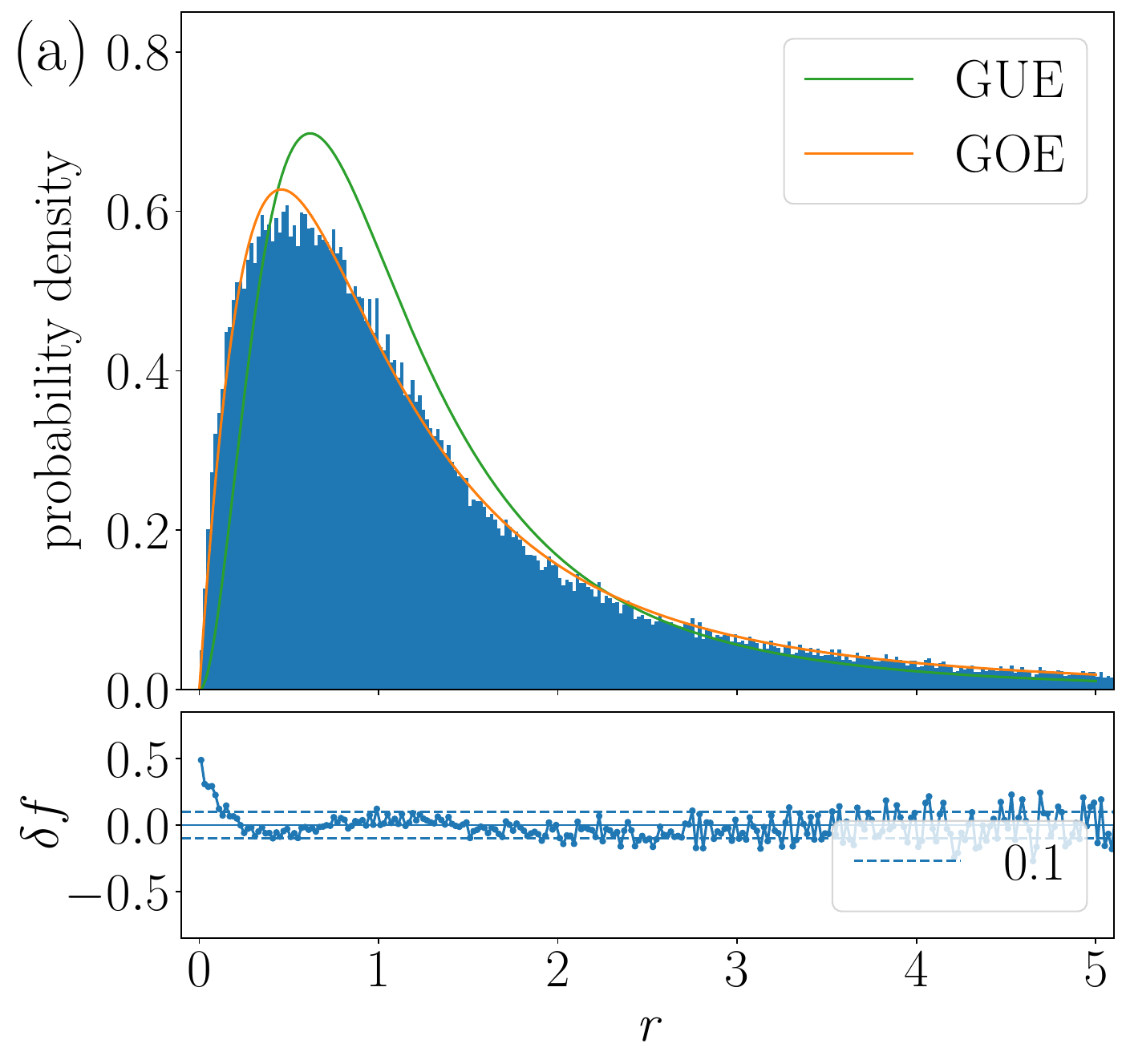}
    \includegraphics[width=0.49\columnwidth]{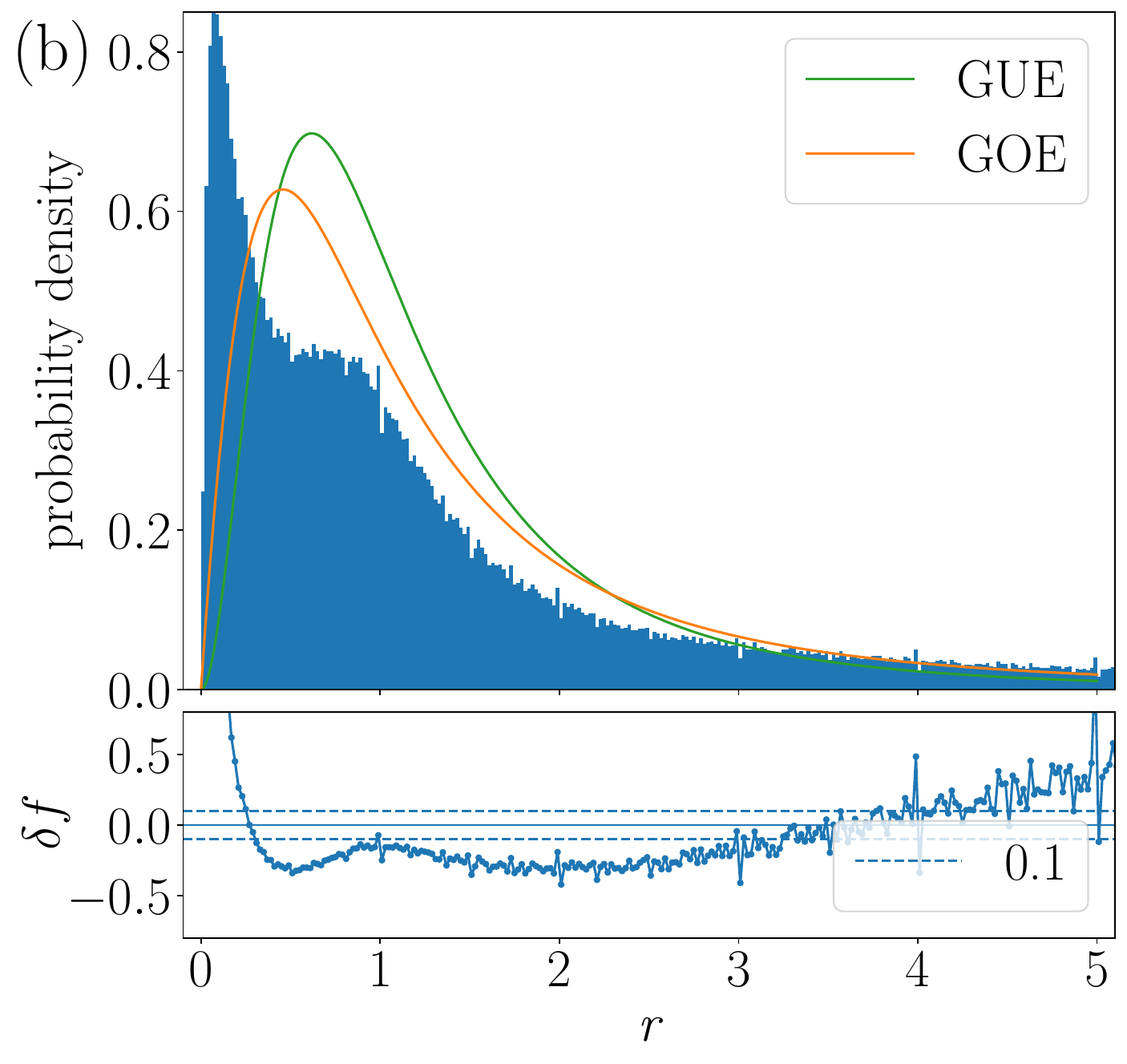}
    \caption{\label{fig:all_stat}Collective statistics for the spacing ratios of extrema of the two-point correlation function after: \textit{(a)}~local operator quench; \textit{(b)}~boundary global quench. The solid lines show analytical GOE (orange) and GUE (green) PDFs. The parameters are: ${L = 1}$, ${m = 10}$, ${\eps = 0.05}$ (for local quench) and ${\tau_0 = 0.01}$ (for global quench). The bottom panels show the deviation between the data and the GOE PDF.}
\end{figure}

The resulting collective statistics are shown in Fig.~\ref{fig:all_stat} in comparison with the PDFs in~\eqref{eq:pdfs}. The values of model parameters ($m$, $L$, and $\eps$) are chosen so that the correlation function clearly develops an irregular pattern. The separation between adjacent slices is ${\Delta x = 0.01}$, the statistical sample is taken up to the cutoff time ${t_f = 500}$, and the number of sampled time points on each slice is ${N_t = 5 \cdot 10^6}$. The statistics do not change significantly when $n_\text{max}$ or $N_t$ is increased further, or when $t_f$ or $\Delta x$ is varied; see Fig.~\ref{fig:robust_loc} in the Supplemental Material.

Fig.~\ref{fig:all_stat}(a) shows that the collective statistics for the local-quench dynamics are consistent with the GOE PDF. The bottom panel shows the deviation $\delta f = f(r)/f_\text{GOE}(r) - 1$; near the maximum of the distribution, the deviation is at the 10\% level. The emergence of GOE, rather than GUE, statistics is consistent with the time-reversal invariance of the system, as in standard level-spacing distribution analyses. The statistics remain close to GOE statistics in a finite window of model parameter values; see Fig.~\ref{fig:dep_loc} in the Supplemental Material.

The distribution for the boundary global quench with values of $m$ and $L$ matching the local-quench case, Fig.~\ref{fig:all_stat}(b), behaves differently and exhibits two features: a narrow structure at small ratios and a bump whose maximum lies close to unity.

\textit{II: Form factors and dip--ramp--plateau structure.---}Beyond level-spacing statistics, one can consider other quantities motivated by RMT as diagnostics of chaotic dynamics~\cite{Haake:2010fgh}. Recently, the spectral form factor (SFF) has received renewed attention~\cite{Cotler:2016fpe, Dyer:2016pou}. In random matrix ensembles, it exhibits a universal late-time behavior known as the dip--ramp--plateau structure~\cite{Liu:2018hlr}. However, recent works show that the interpretation of the SFF can be subtle: ramp-like structures may arise even when the level-spacing statistics are not described by standard RMT ensembles~\cite{Das:2022evy, Das:2023yfj, DeClerck:2023fax, Jeong:2024jjn, Ageev:2024gem}.

The SFF encodes pairwise correlations between levels in the spectrum~$\{\lambda_i\}$. Specifically, the SFF is defined by the Fourier transform of a two-level correlation function~\cite{Liu:2018hlr} which yields
\be
    g(t) = g_0\sum_{i,\,j\,=\,1}^{N} e^{-i(\lambda_i - \lambda_j)t},
    \label{eq:SFF_spectrum}
\ee
where $N$~is the number of levels taken into account and $g_0$~is a normalization constant, chosen so that ${g(0) = 1}$. 

For quantum field theories, one encounters the same obstacle as in level-spacing distribution analyses: the calculation requires explicit knowledge of the system's spectrum. In our setup, we consider two alternatives: a modified form factor that captures the spacetime structure of correlations, and the spectral form factor computed from the partition function. The latter is the conventional SFF, but expressed in terms of the partition function, an object naturally suited to quantum field theory. One may also propose other state-dependent versions of the form factor; for an example, see Fig.~\ref{fig:FF_modes} in the Supplemental Material.

\textit{Spacetime correlation form factor.} Ref.~\cite{Bianchi:2024fsi} proposed a peak-distribution-based form factor construction, in which the energy levels $\{\lambda_i\}$ in the definition~\eqref{eq:SFF_spectrum} are replaced by the unfolded extrema positions $\{\tilde{z}_i\}$ with the corresponding Fourier-conjugate variable $s$. In string-scattering problems, this quantity is called the scattering form factor. Here, in the absence of scattering, we call it the spacetime correlation form factor.

In order to extract the universal correlations encoded in the form factor, the unfolding procedure that produces a constant mean density across the data set is essential~\cite{Bianchi:2024fsi}. Given the raw extrema positions $z$, the unfolded variable $\tilde{z}$ is defined as
\be
    \tilde{z} = I(z), \quad I(z) = N_\text{ext}\int\limits_0^z\rho(z')dz',
\ee
where $I(z)$ is the cumulative density, $\rho(z)$ is the density of extrema positions, and $N_\text{ext}$ is the number of extrema in the set. Note that we construct a separate unfolding map for each slice at fixed spatial coordinate; we use the same set of slices as in the construction of the spacing-ratio statistics of extrema.

\begin{figure}[t]
    \includegraphics[width=0.45\columnwidth]{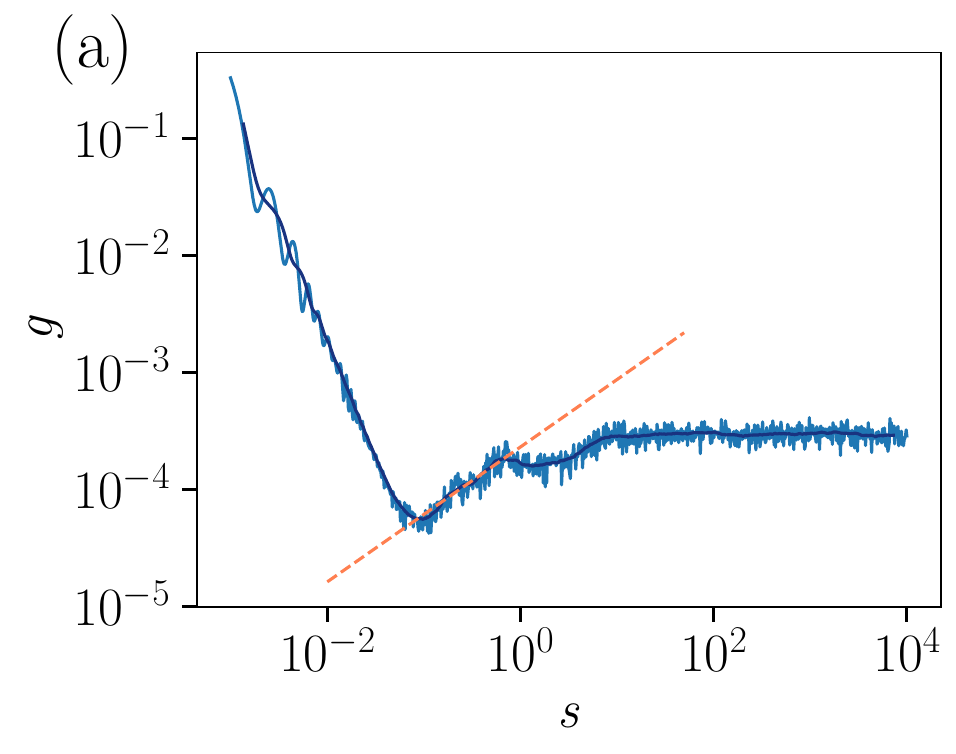}
    \includegraphics[width=0.45\columnwidth]{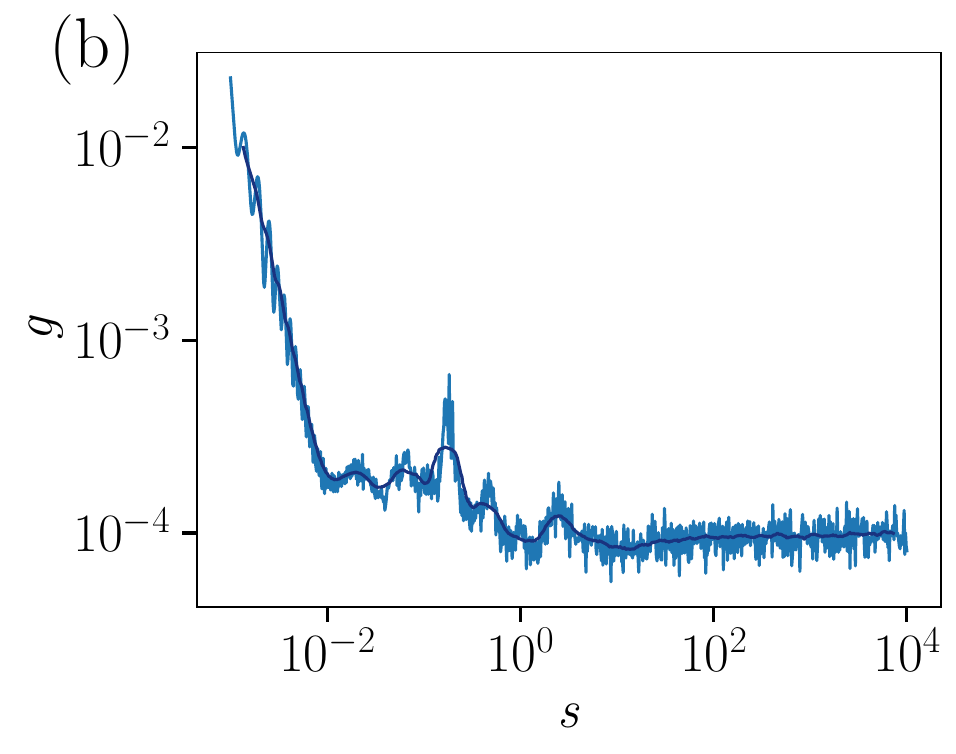}
    \caption{\label{fig:scFF}Spacetime correlation form factor for: \textit{(a)}~local operator quench; \textit{(b)}~boundary global quench. The parameters are: ${L = 1}$, ${m = 10}$, ${\eps = 0.05}$ (for local quench) and ${\tau_0 = 0.01}$ (for global quench). The darker line shows the moving average of the data; the dashed orange line shows a linear fit to the averaged curve over the ramp region.}
\end{figure}

After unfolding and averaging over slices, the spacetime correlation form factor in the local-quench case displays a dip--ramp--plateau structure for the same model parameters used to analyze the spacing-ratio statistics, as shown in Fig.~\ref{fig:scFF}. By contrast, the global-quench case does not exhibit a clear ramp.

\begin{figure}[t]
    \includegraphics[width=0.45\columnwidth]{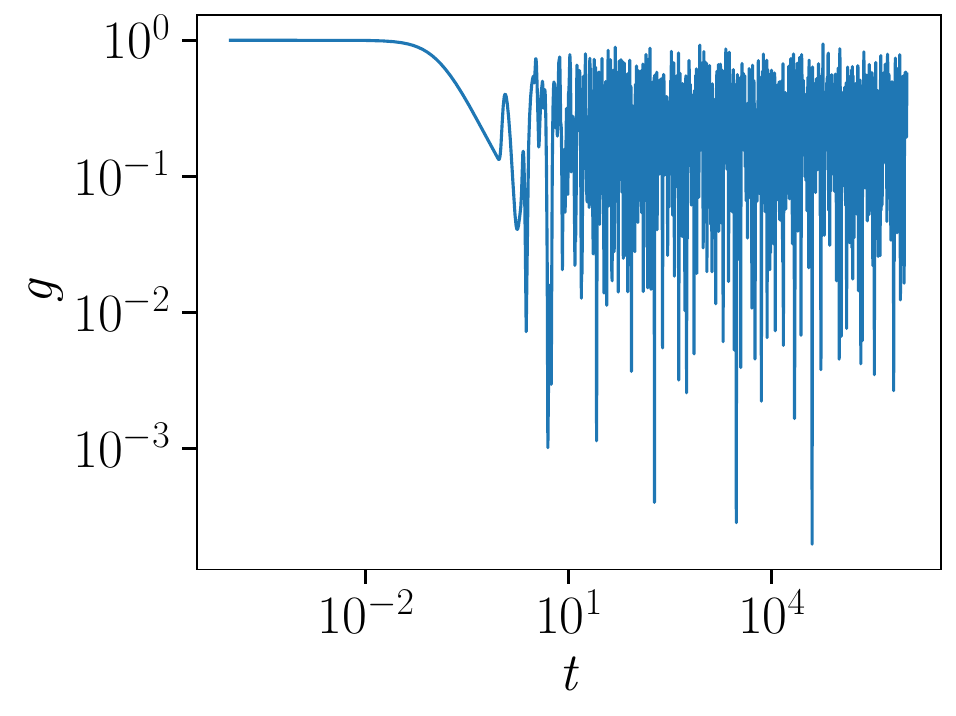}
    \caption{\label{fig:SFF} Spectral form factor calculated from the partition function. The parameters are: ${L = 1}$, ${m = 10}$, and ${\eps = 0.05}$.}
\end{figure}

\textit{SFF from the partition function.} Besides~\eqref{eq:SFF_spectrum}, the SFF can also be defined in terms of the analytically-continued partition function~$Z$ for a system at inverse finite temperature~$\beta$~\cite{Cotler:2016fpe}, 
\be
    g(\beta, t) = \left|\frac{Z(\beta + it)}{Z(\beta)}\right|^2.
    \label{eq:SFF}
\ee
The partition function here is the thermal partition function, $Z(\beta) = \text{Tr}\,e^{-\beta H}$. More generally, for a density matrix $\rho$, not necessarily thermal, we define $Z(t) = \text{Tr}(e^{-iHt}\rho)$. In our local-quench setup, $\rho = \ket{\Psi}\bra{\Psi}$; using the mode decomposition, the SFF evaluates to
\be
    g(t) = g_0\sum_n\sum_s p_n p_s e^{-i(\om_n - \om_s)t},
\ee
where $p_n = e^{-2\eps\om_n}/(2\om_n)$ and $g_0 = (\sum_n p_n)^{-2}$. This quantity shows a clear plateau; however, no ramp is observed, as shown in \figref{fig:SFF}.

\textit{Conclusions.---}We studied a massive free scalar field theory in a finite volume and showed that driving this system locally out of equilibrium by a local operator quench can induce RMT-like universal behavior in observable statistics. The absence of a ramp in the standard SFF is consistent with the underlying free spectrum; nevertheless, the observable statistics based on two-point correlation function extrema mimic features found in chaotic systems, such as those in highly excited string scattering amplitudes.

The relevant universal object in this setting is not the fluctuations of the spectrum, but the spacetime correlation structure. The local operator insertion in the large-mass regime excites broadly dispersed finite-volume modes; their propagation around the compact spatial dimension and superposition at different spacetime points generate irregular patterns.

The GOE-like observable statistics emerge within a finite window of parameter space in which this irregular spacetime structure is well developed. The crossover to more regular behavior outside this regime suggests that the effect requires a balance between locality, broad mode excitation, and finite-volume winding.

It would be interesting to extend this analysis to higher-dimensional models, interacting theories and other quench protocols.

\begin{acknowledgments}
    \textit{Acknowledgments.---}We would like to thank Jacob Sonnenschein for useful comments and especially for the advice to analyze the spectral form factor. This work was supported by the Russian Science Foundation under grant No.\,24-72-10061, \href{https://rscf.ru/en/project/24-72-10061/}{https://rscf.ru/en/project/24-72-10061/}.
\end{acknowledgments}

\appendix
\section{Supplemental Material}

\begin{figure}[ht]
    \includegraphics[width=0.45\columnwidth]{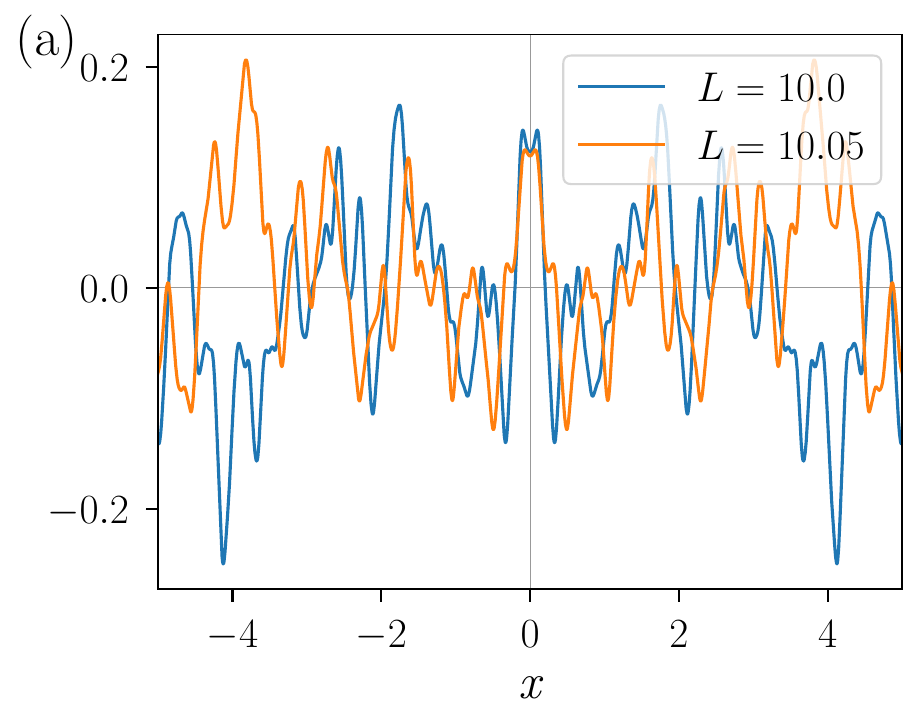}
    \includegraphics[width=0.45\columnwidth]{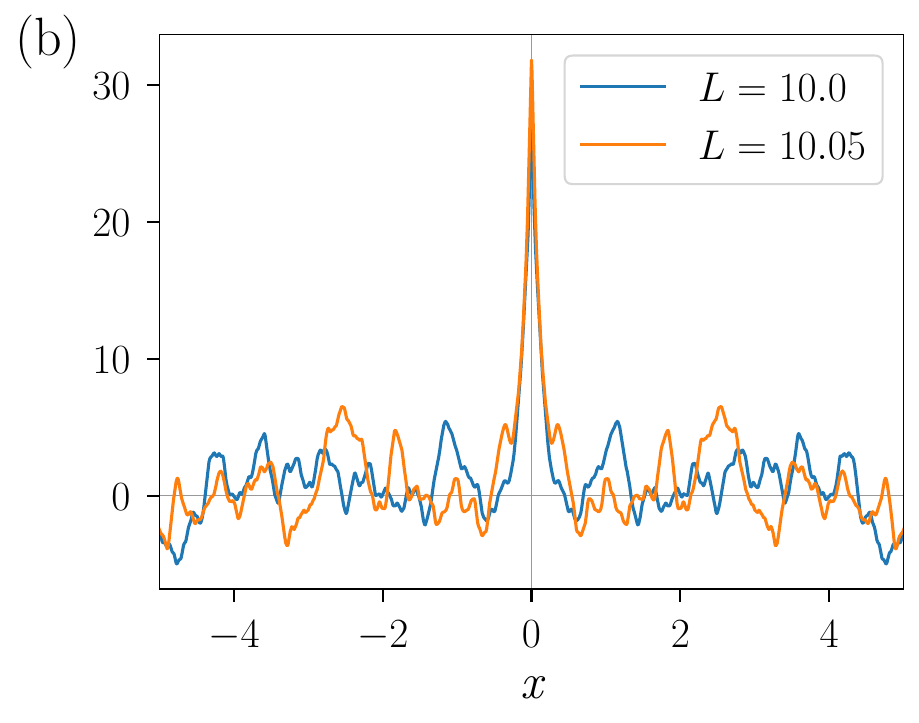}
    \caption{\label{fig:profiles_L}Two-point correlation function with small variations in the cylinder circumference~$L$ at a fixed large time moment (${t = 500}$) after: \textit{(a)}~local operator quench; \textit{(b)}~boundary global quench. The parameters are: ${m = 10}$, ${L = 10}$ (blue line) and ${L = 10.05}$ (orange line), ${\eps = 0.05}$ (for local quench) and ${\tau_0 = 0.01}$ (for global quench).}
\end{figure}

\textit{Two-point function dependence on the circumference.---}The sample profiles of the nonequilibrium two-point function are as sensitive to small variations in the cylinder circumference as to those in the mass (see \figref{fig:profiles_L}): even small changes lead to significantly different profiles.

\begin{figure}[ht]
    \includegraphics[width=0.45\columnwidth]{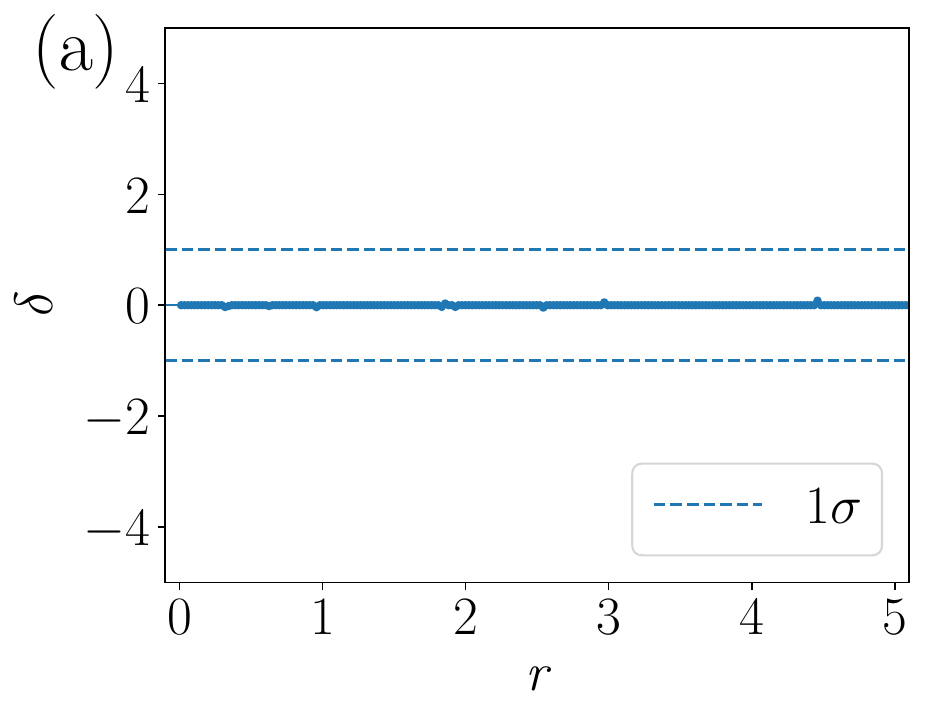}
    \includegraphics[width=0.45\columnwidth]{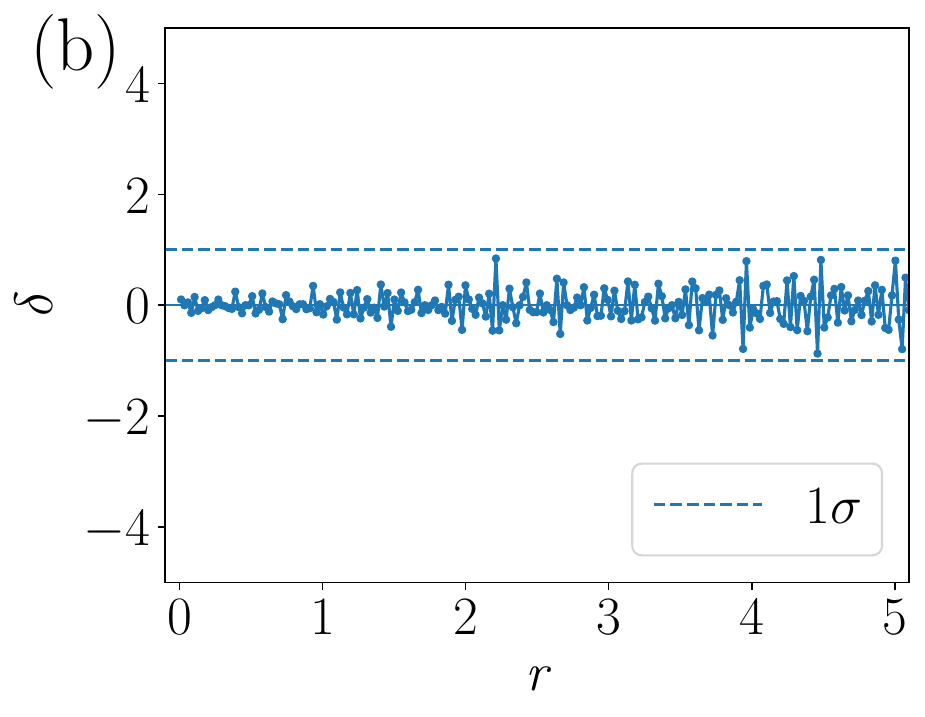}\\
    \includegraphics[width=0.45\columnwidth]{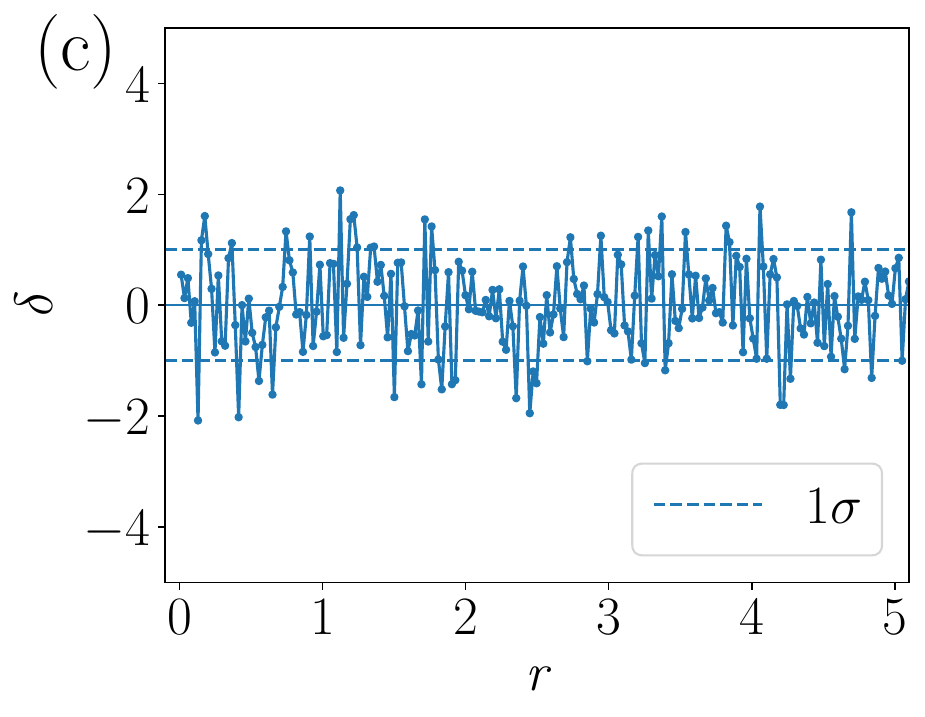}
    \includegraphics[width=0.45\columnwidth]{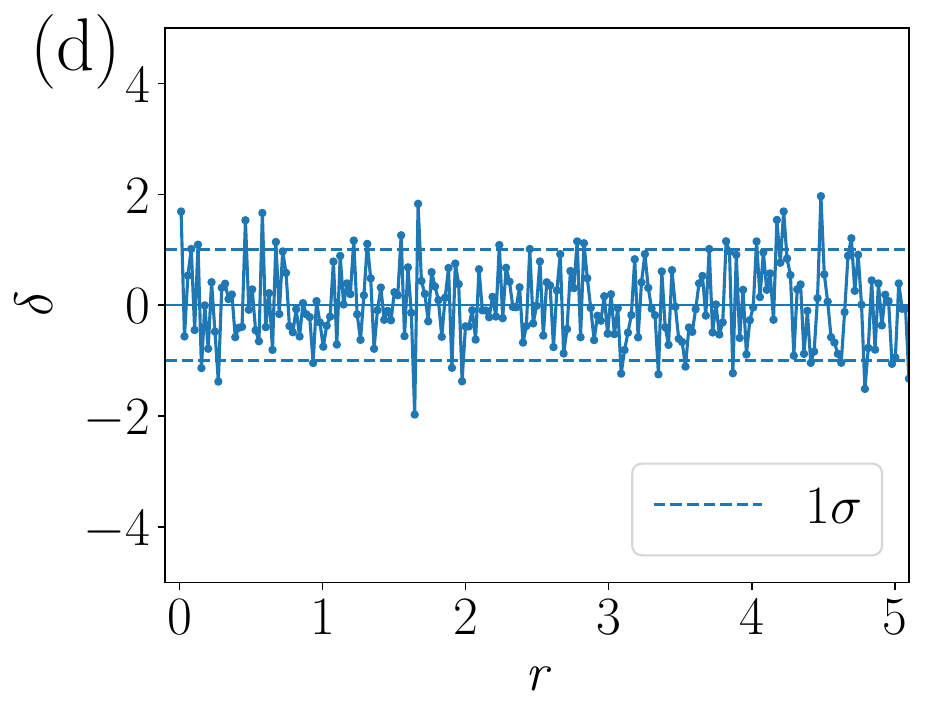}
    \caption{\label{fig:robust_loc}Residuals between the distributions of the collective statistics in the local-quench setup, comparing: \textit{(a)}~${N = 250}$ and ${N = 750}$; \textit{(b)}~${N_t = 5\cdot10^6}$ and ${N_t = 15\cdot10^6}$; \textit{(c)}~${t_f = 500}$ with ${N_t = 5\cdot10^6}$ and ${t_f = 1500}$ with ${N_t = 15\cdot10^6}$; \textit{(d)}~${\Delta x = 0.01}$ and ${\Delta x = 0.1}$. The default values of the parameters are: ${N = 250}$, ${N_t = 5\cdot10^6}$, ${t_f = 500}$, ${\Delta x = 0.01}$, ${L = 1}$, ${m = 10}$, and ${\eps = 0.05}$.}
\end{figure}

\begin{figure}[ht]
    \includegraphics[width=0.45\columnwidth]{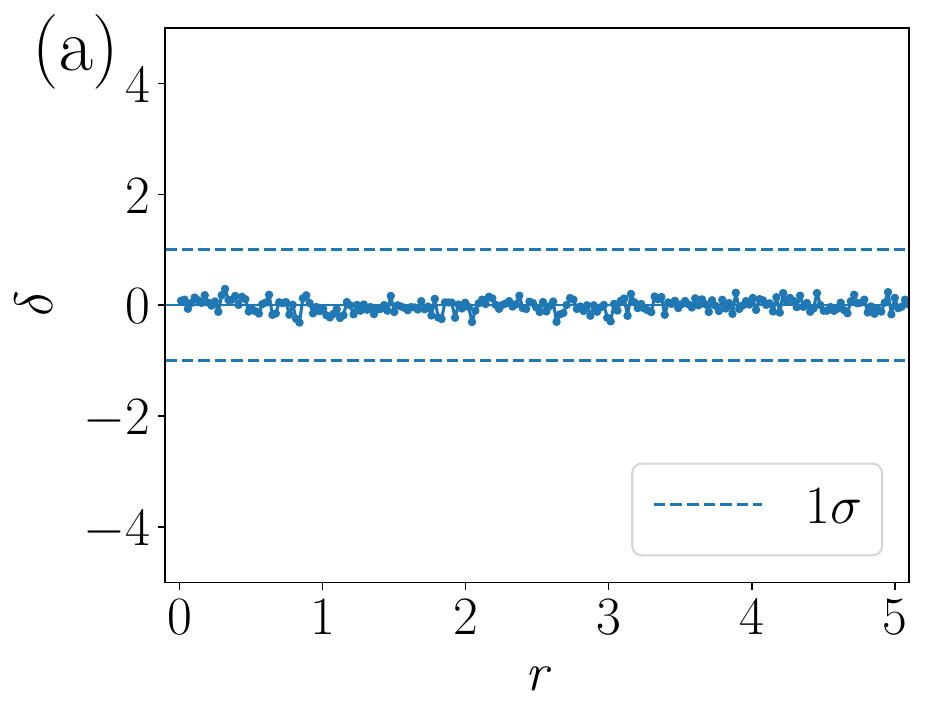}
    \includegraphics[width=0.45\columnwidth]{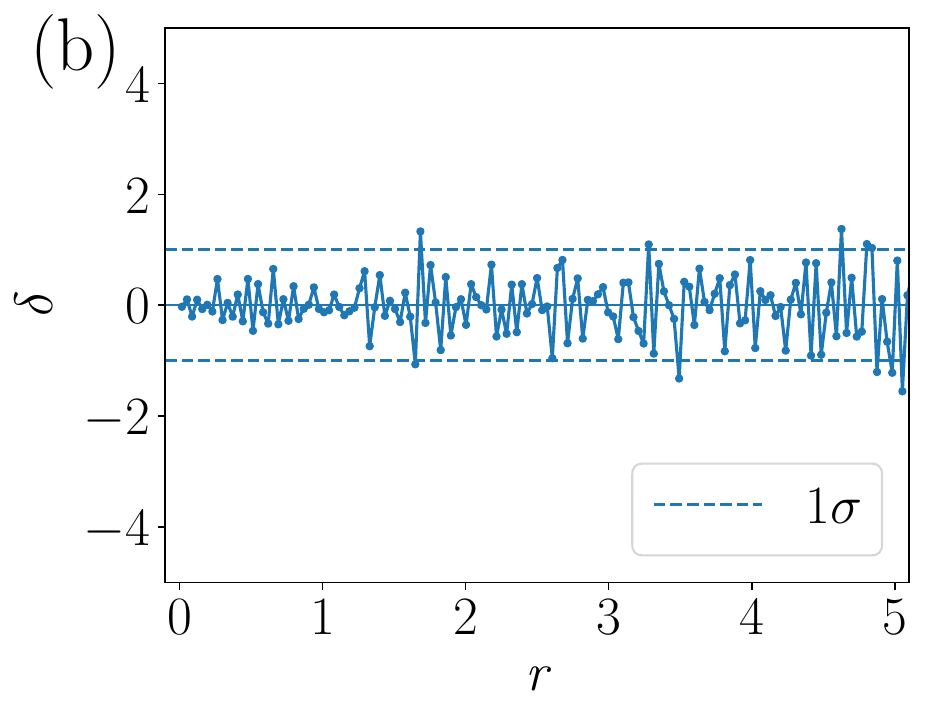}\\
    \includegraphics[width=0.45\columnwidth]{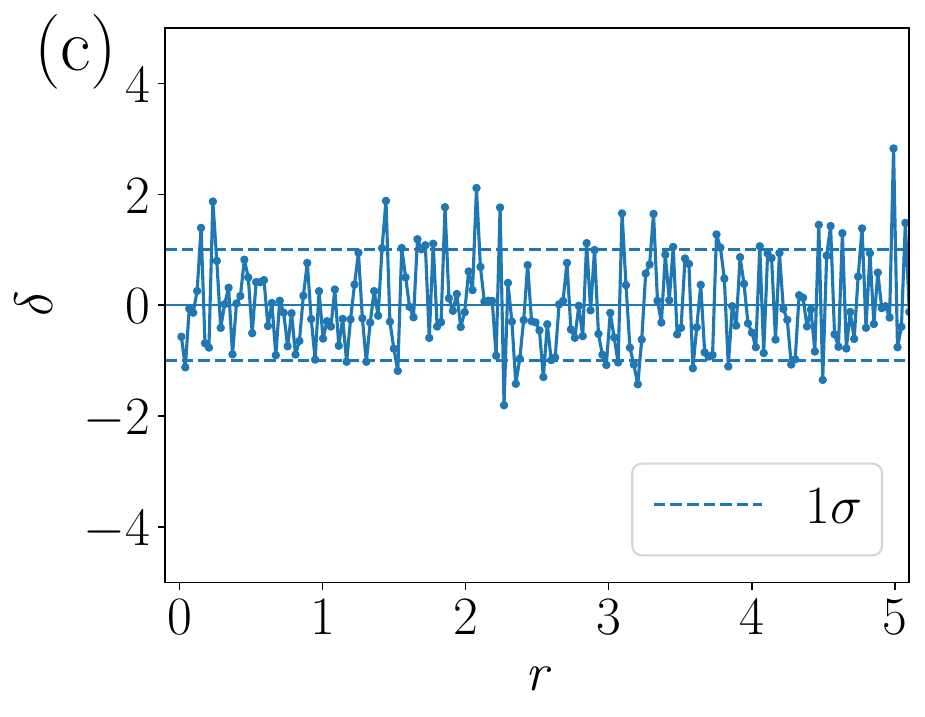}
    \includegraphics[width=0.45\columnwidth]{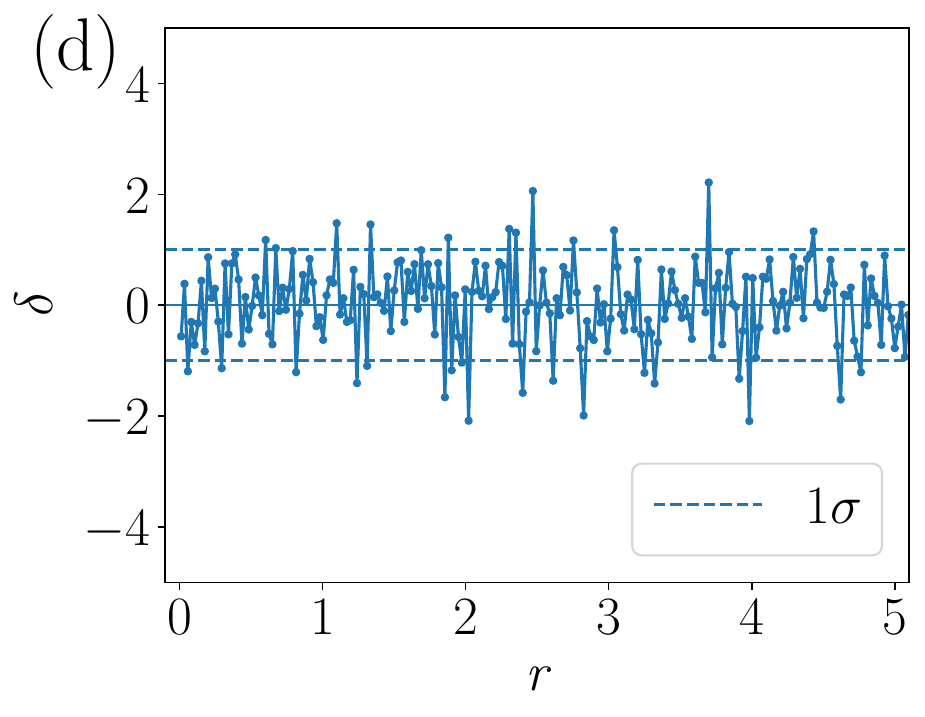}
    \caption{\label{fig:robust_glob}Residuals between the distributions of the collective statistics in the global-quench setup, comparing: \textit{(a)}~${N = 250}$ and ${N = 750}$; \textit{(b)}~${N_t = 5\cdot10^6}$ and ${N_t = 15\cdot10^6}$; \textit{(c)}~${t_f = 500}$ with ${N_t = 5\cdot10^6}$ and ${t_f = 1500}$ with ${N_t = 15\cdot10^6}$; \textit{(d)}~${\Delta x = 0.01}$ and ${\Delta x = 0.03}$. The default values of the parameters are: ${N = 250}$, ${N_t = 5\cdot10^6}$, ${t_f = 500}$, ${\Delta x = 0.01}$, ${L = 1}$, ${m = 10}$, and ${\tau = 0.01}$.}
\end{figure}

\textit{Robustness of the statistics.---}The collective statistics for the spacing ratios of extrema of the two-point correlation function are stable with respect to changes in numerical calculation parameters (the number of series terms $N$ and the number of points along the slices at fixed spatial coordinate $N_t$) as well as to increasing the sample size by extending the cutoff time $t_f$ and by decreasing the separation between slices $\Delta x$. The residuals between pairs of probability densities are shown in \figref{fig:robust_loc} for the local quench and in \figref{fig:robust_glob} for the global quench; the residuals are defined as
\be
    \delta = \frac{p_1 - p_2}{\sqrt{\sigma_1^2 + \sigma_2^2}},
\ee
where $p_{1,2}$ are the binned probability densities and $\sigma_{1,2}$ are the corresponding standard deviations in each bin. In the region near the maxima of the probability densities, the statistics remain close to those for the default choice of parameters up to random noise with mean approximately zero and residuals lying within a $2\sigma$ band.

\begin{figure*}
    \includegraphics[width=0.45\columnwidth]{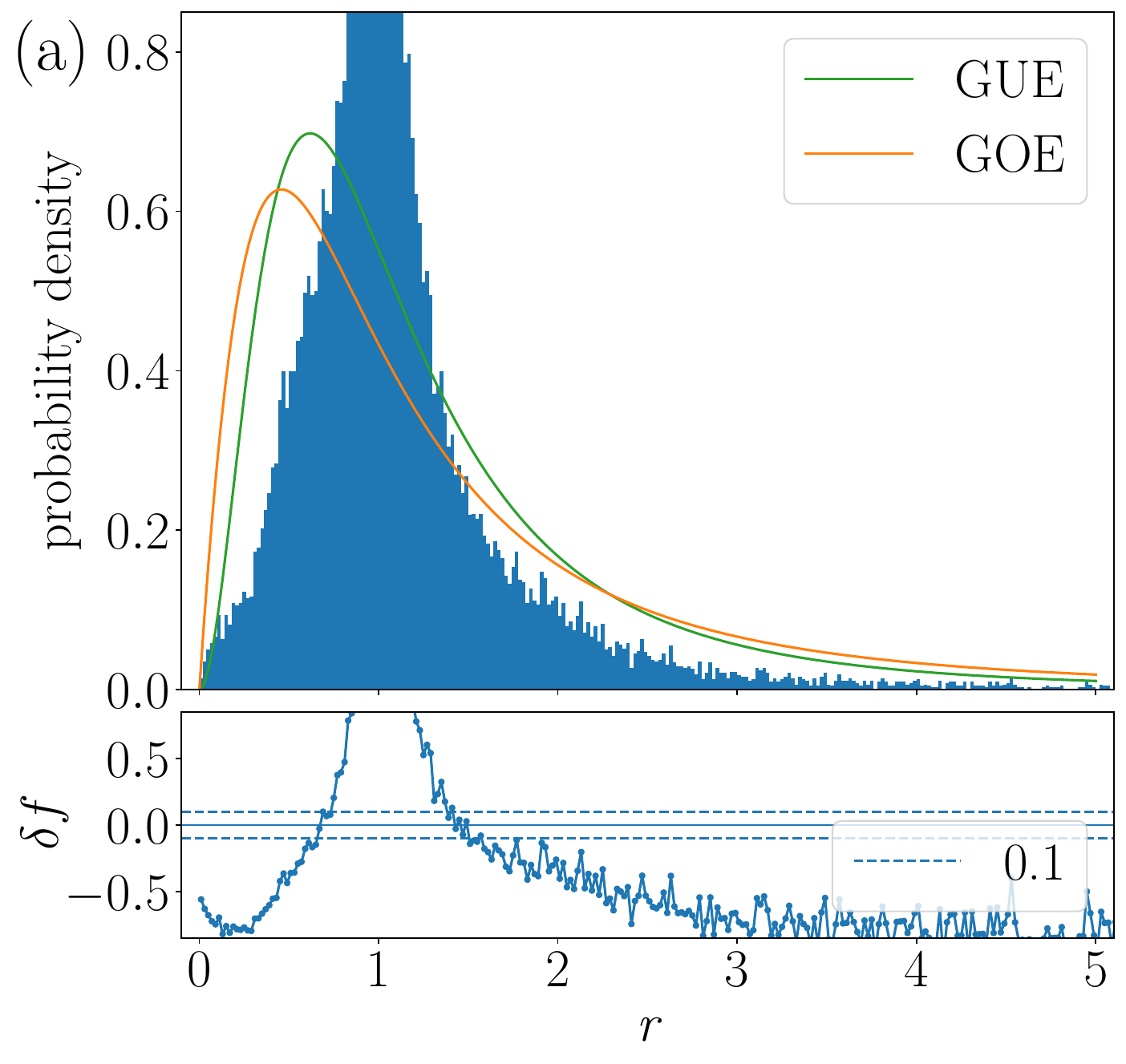}
    \includegraphics[width=0.45\columnwidth]{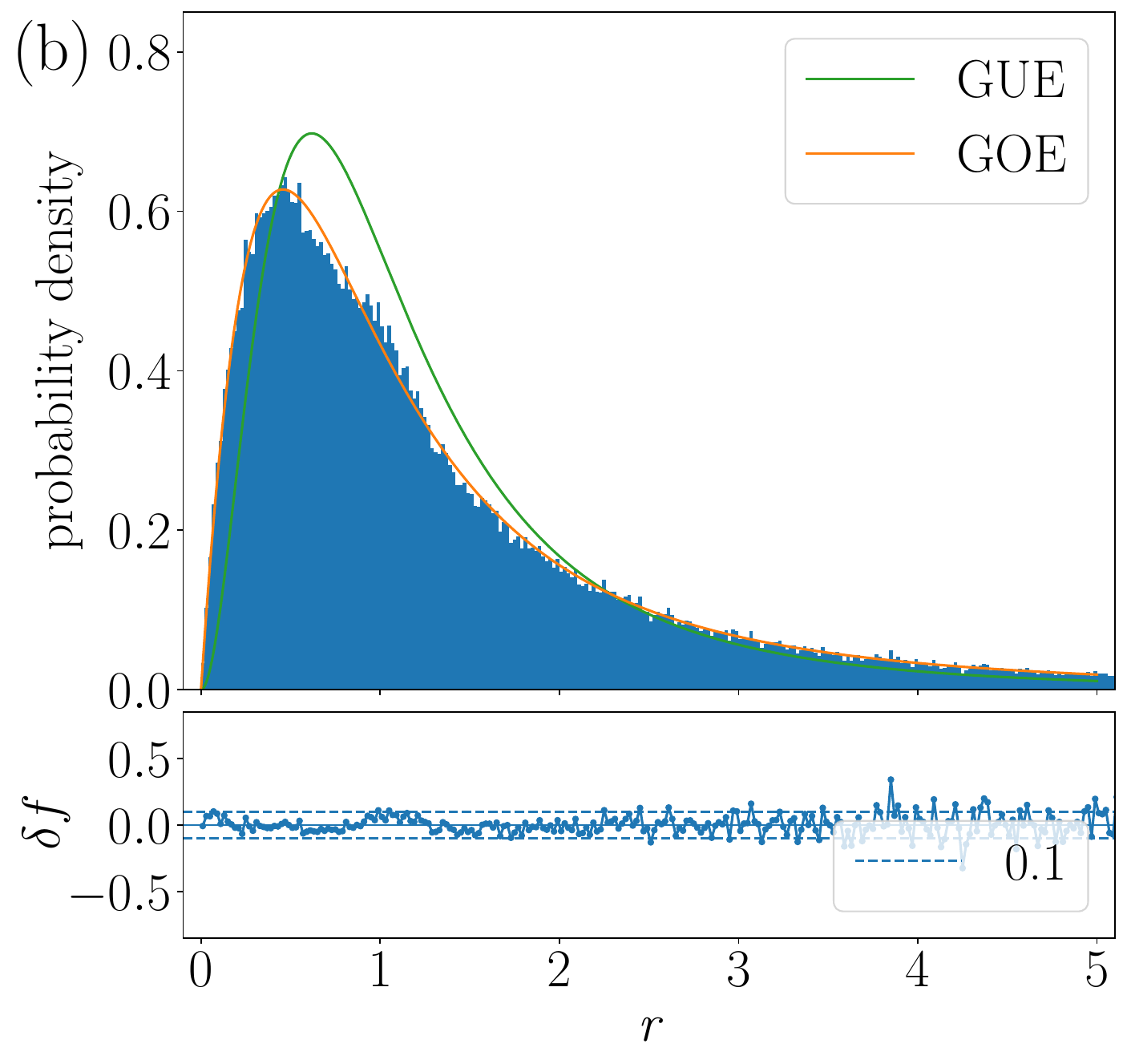}
    \includegraphics[width=0.45\columnwidth]{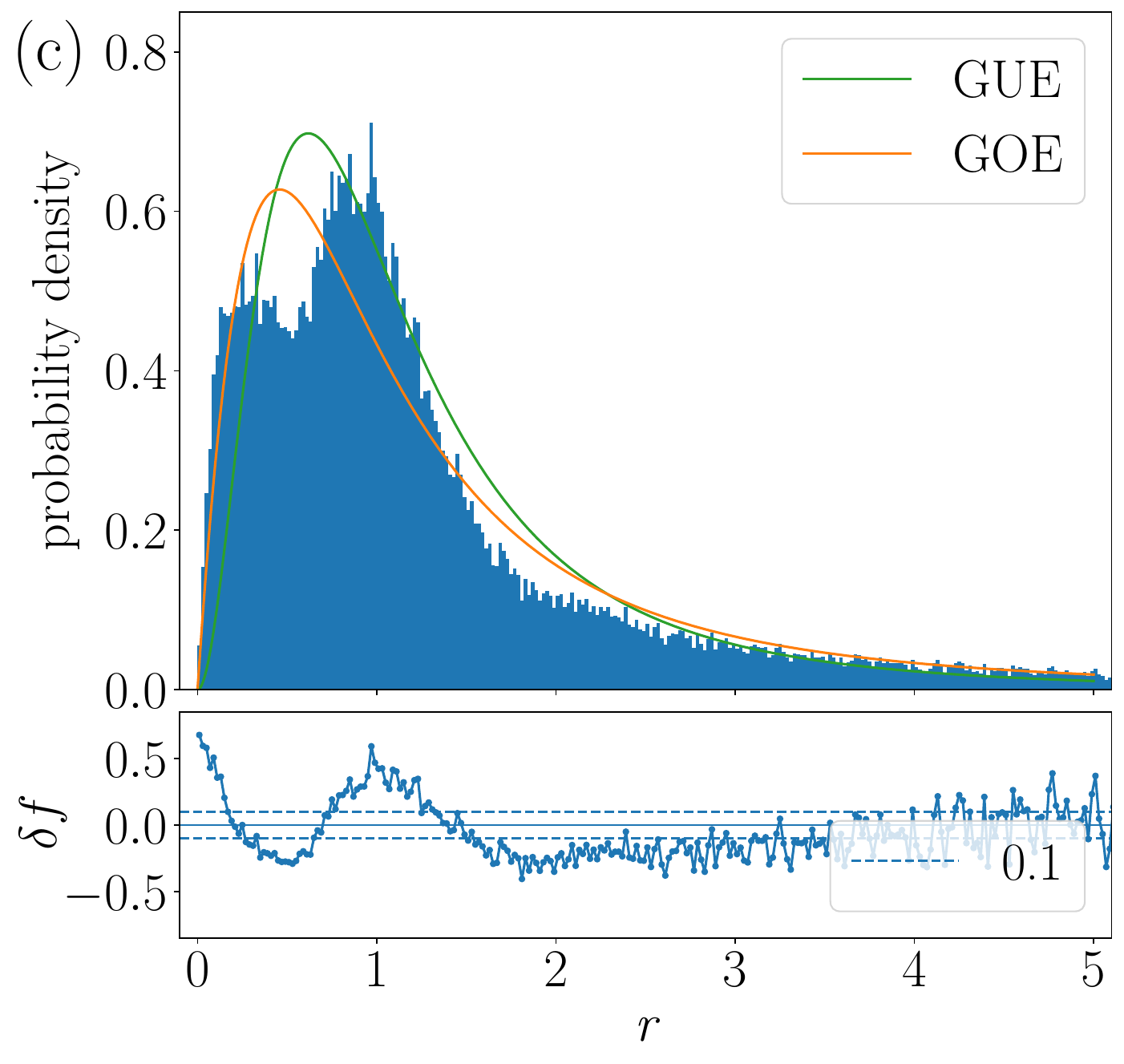}\\
    \includegraphics[width=0.45\columnwidth]{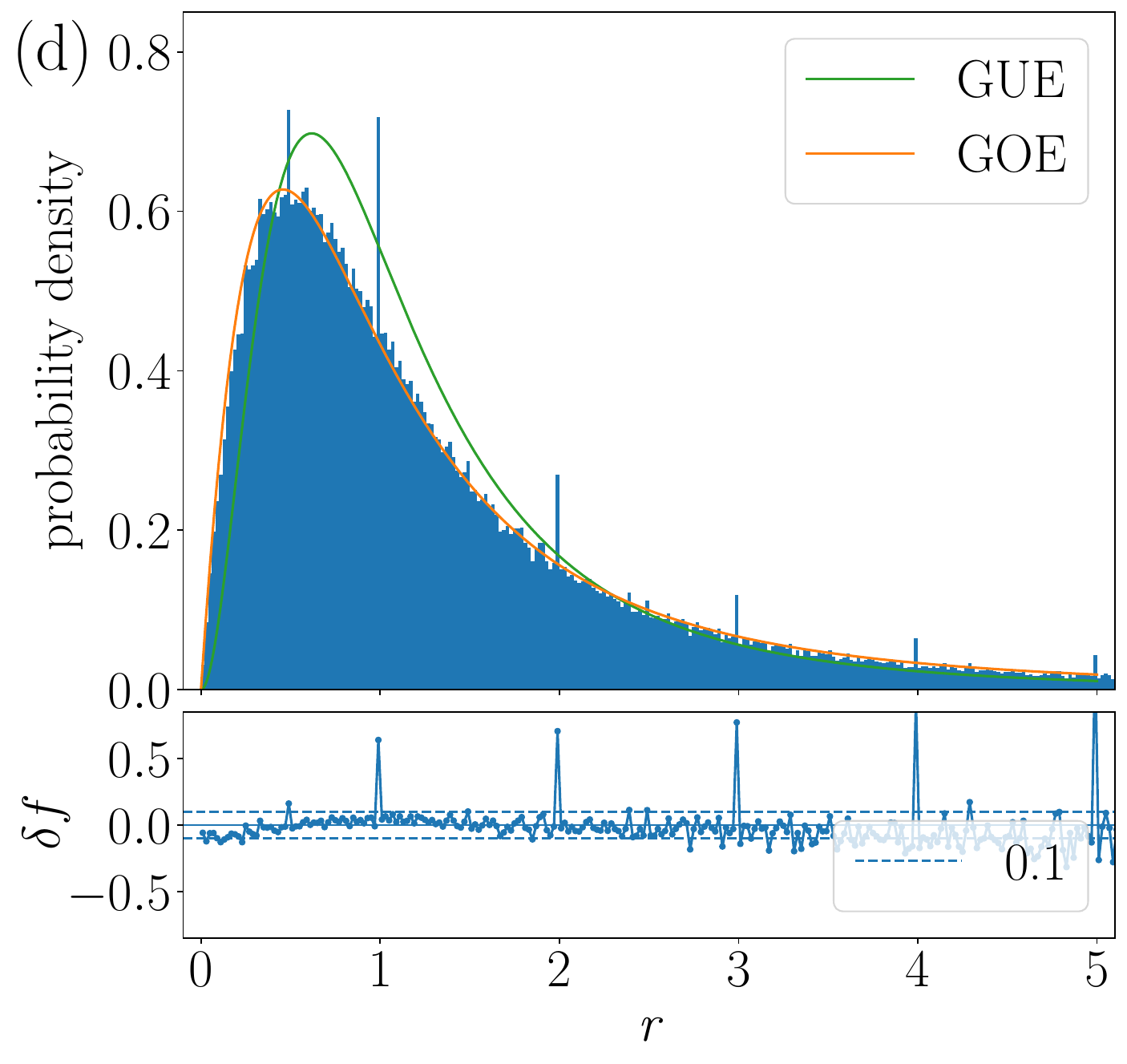}
    \includegraphics[width=0.45\columnwidth]{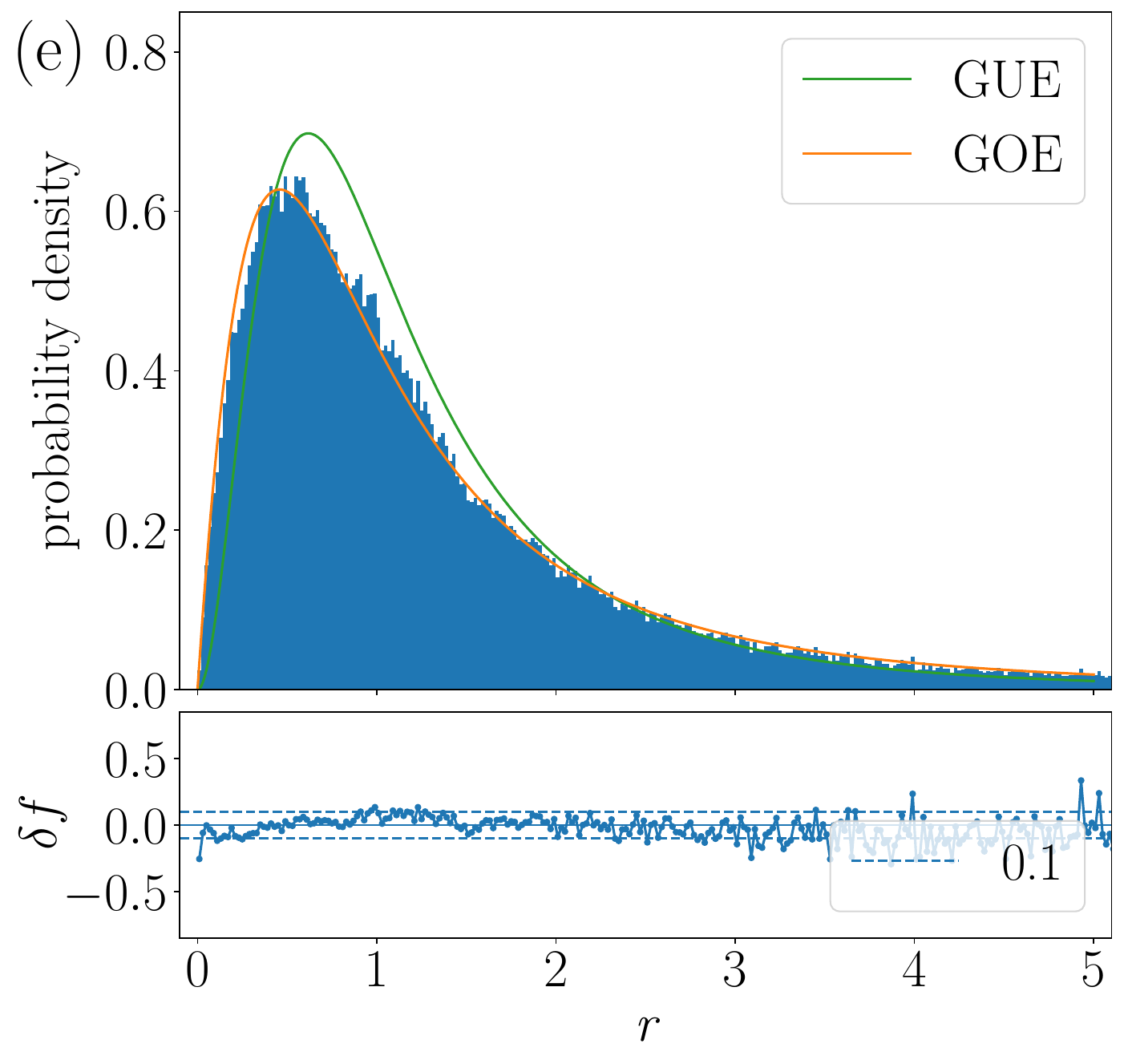}
    \includegraphics[width=0.45\columnwidth]{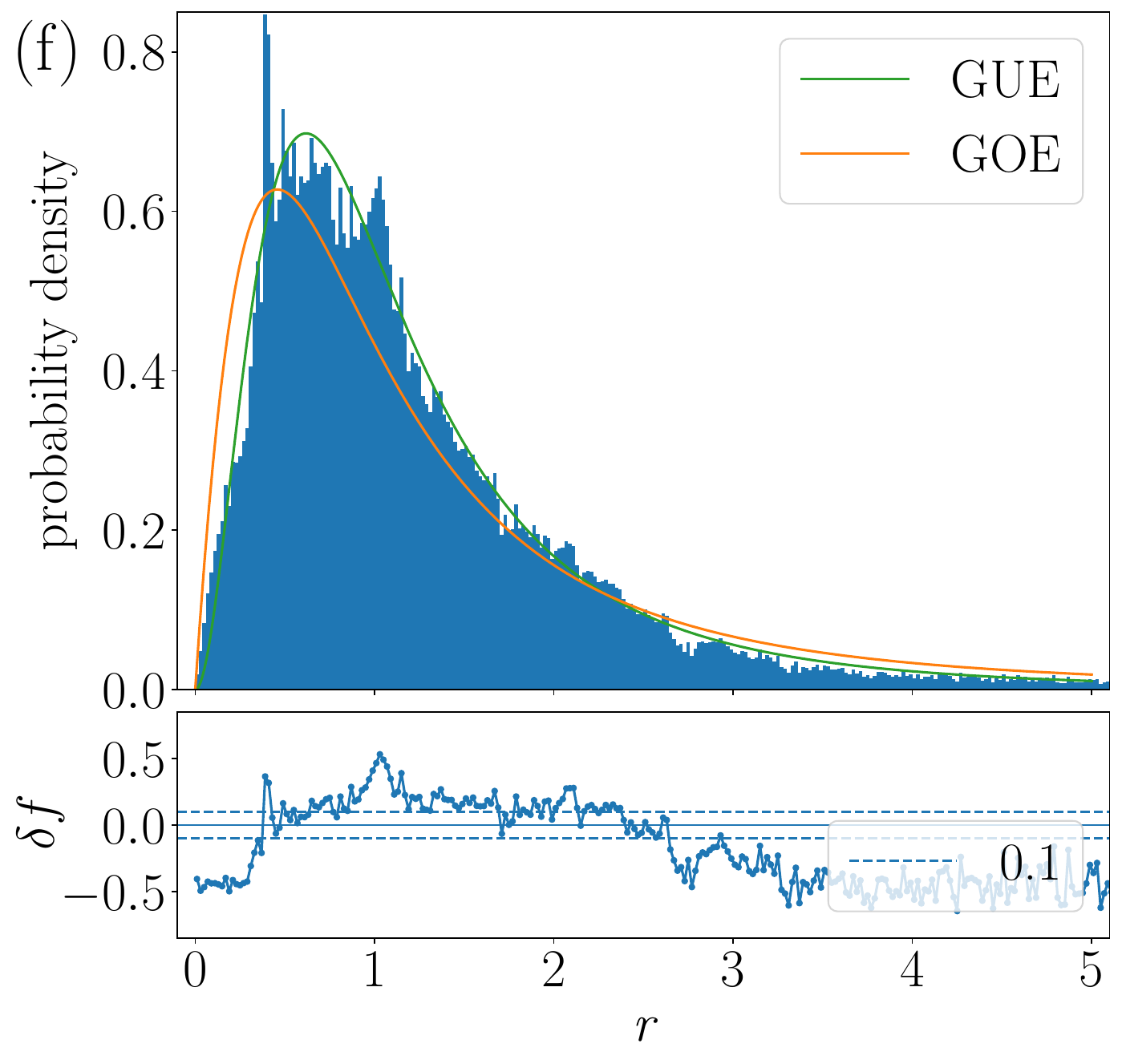}\\
    \includegraphics[width=0.45\columnwidth]{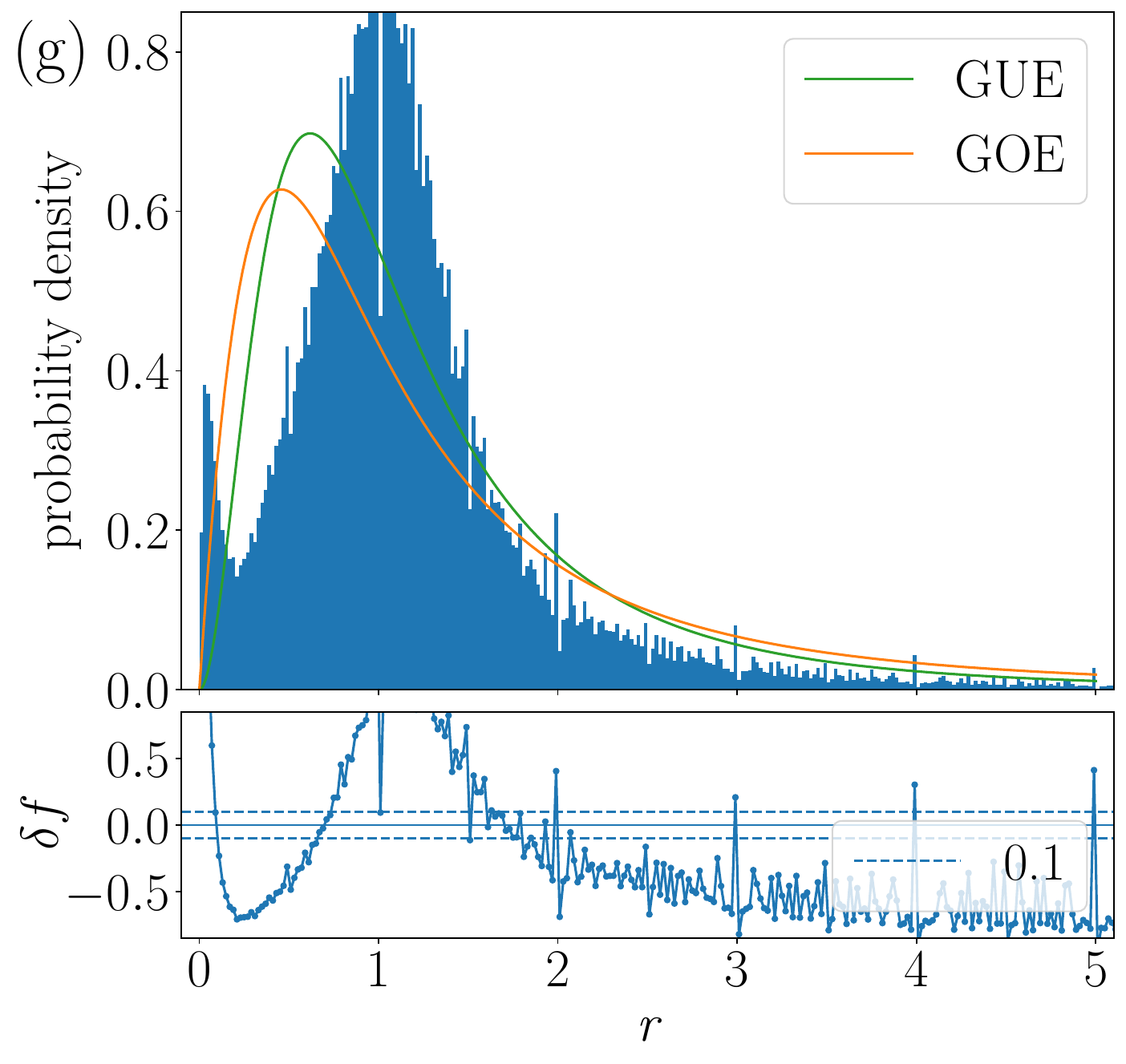}
    \includegraphics[width=0.45\columnwidth]{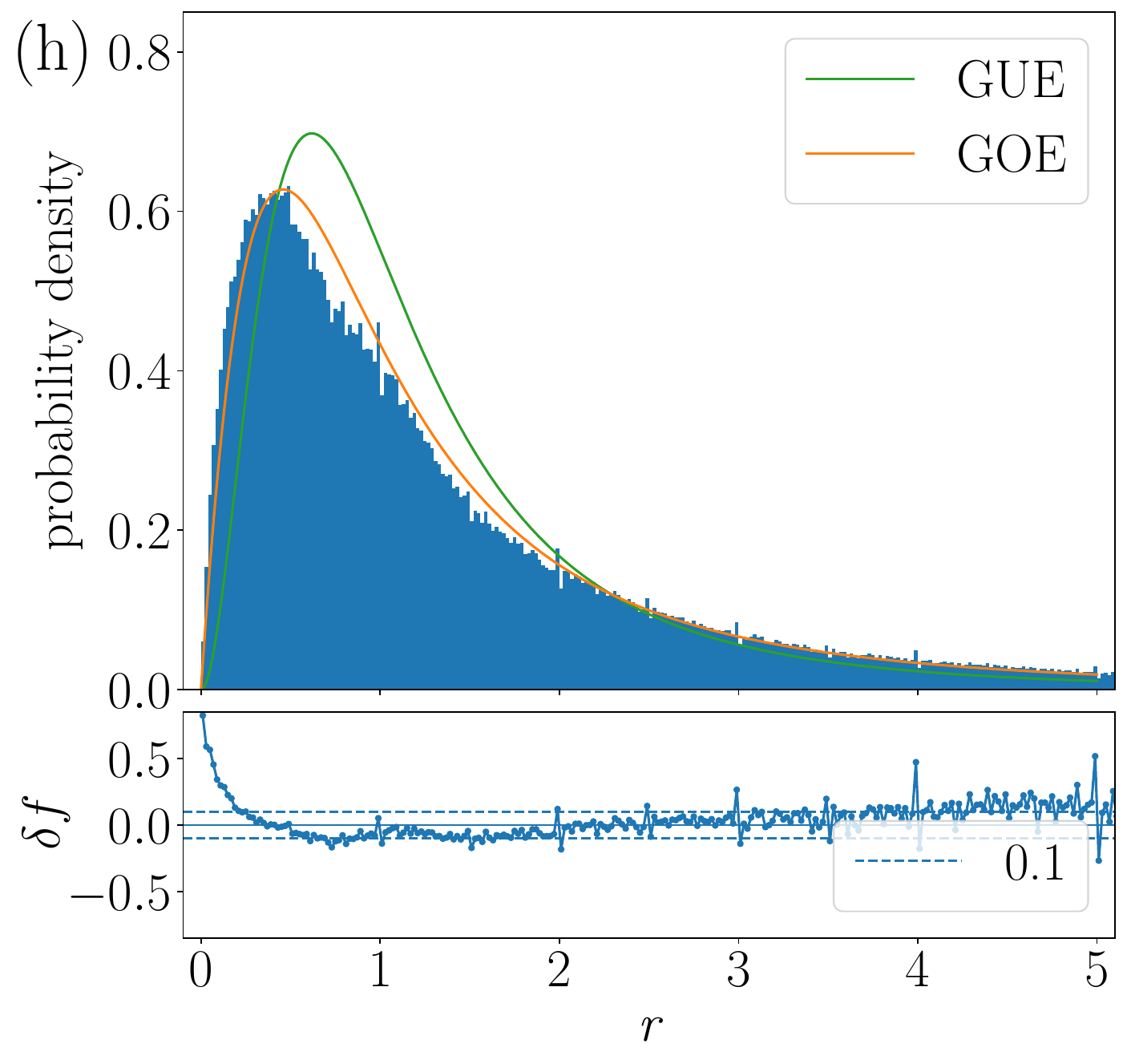}
    \includegraphics[width=0.45\columnwidth]{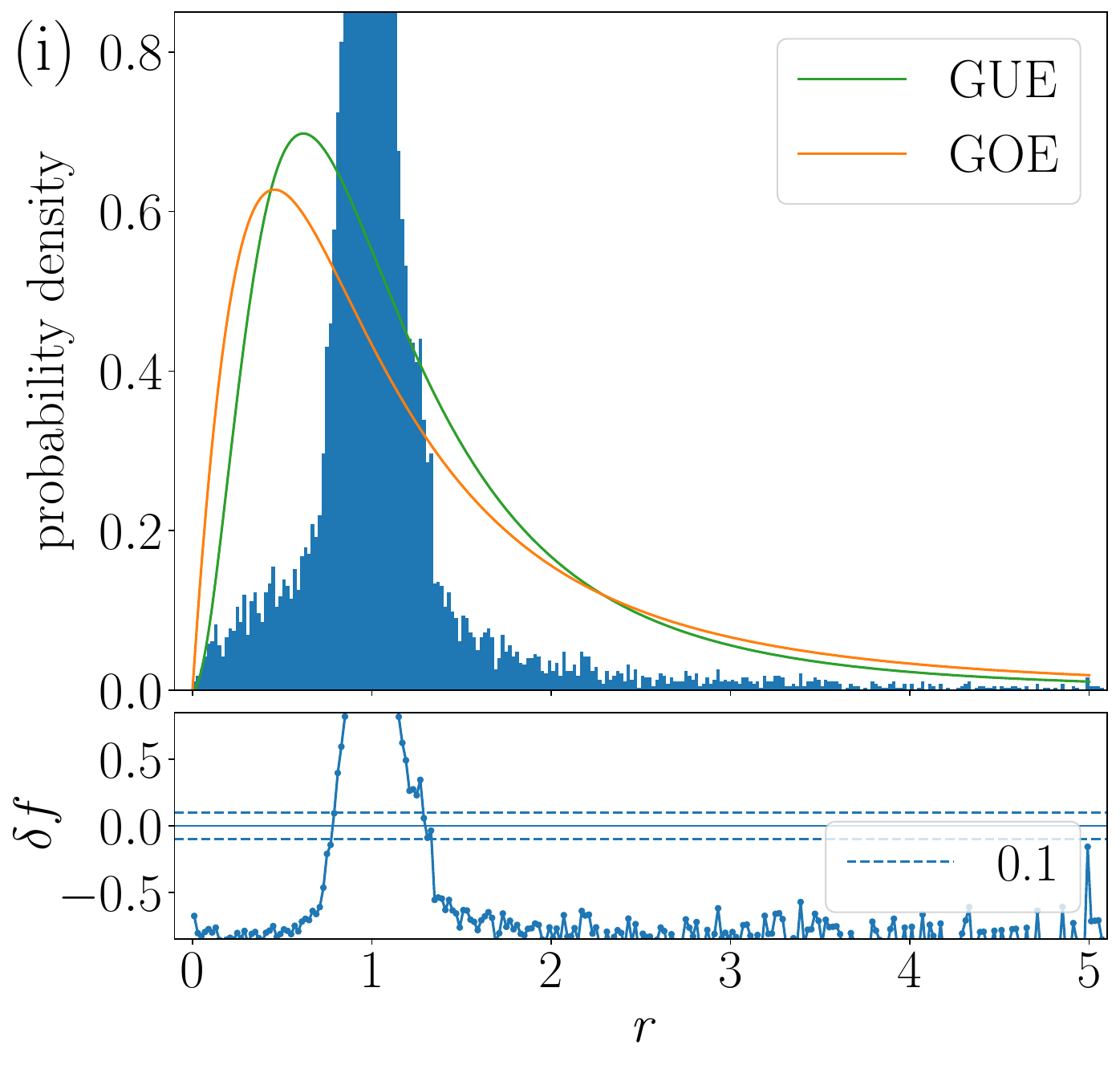}\\
    \caption{\label{fig:dep_loc}Collective statistics of the spacing ratios of extrema in the local-quench setup for parameter values different from the default values: \textit{(a)}~larger circumference, $L = 100$ ($\Delta x = 1$);  \textit{(b)}~larger circumference, $L = 10$ ($\Delta x = 0.1$); \textit{(c)}~smaller circumference, $L = 0.5$; \textit{(d)}~larger mass, ${m = 100}$; \textit{(e)}~larger mass, ${m = 50}$; \textit{(f)}~smaller mass, ${m = 0.1}$; \textit{(g)}~smaller UV-regulator, ${\eps = 0.001}$; \textit{(h)}~smaller UV-regulator, ${\eps = 0.01}$; \textit{(i)}~larger UV-regulator, ${\eps = 0.5}$. The default values of the parameters are: ${N = 250}$, ${N_t = 5\cdot10^6}$, ${t_f = 500}$, ${\Delta x = 0.01}$, ${L = 1}$, ${m = 10}$, and ${\eps = 0.05}$. The bottom panels show the relative deviation between the data and the GOE PDF.}
\end{figure*}

\textit{Dependence of the statistics on model parameters.---}The model parameters that influence the correlation function pattern after the local operator quench are the mass $m$, the compact dimension circumference $L$, and the UV-regulator $\eps$. The default values, $m = 10$, $L = 1$, and $\eps = 0.05$ are chosen so that the pattern is clearly erratic. In \figref{fig:dep_loc}, the distributions obtained for values larger and smaller than the default are compared. Statistics close to the GOE statistics emerge within a finite range of model parameters with substantial stability under increases in the mass. Outside this window, crossovers back to more regular regimes characterized by a distribution peaked near unity occur.

\begin{figure*}
    \includegraphics[width=0.45\columnwidth]{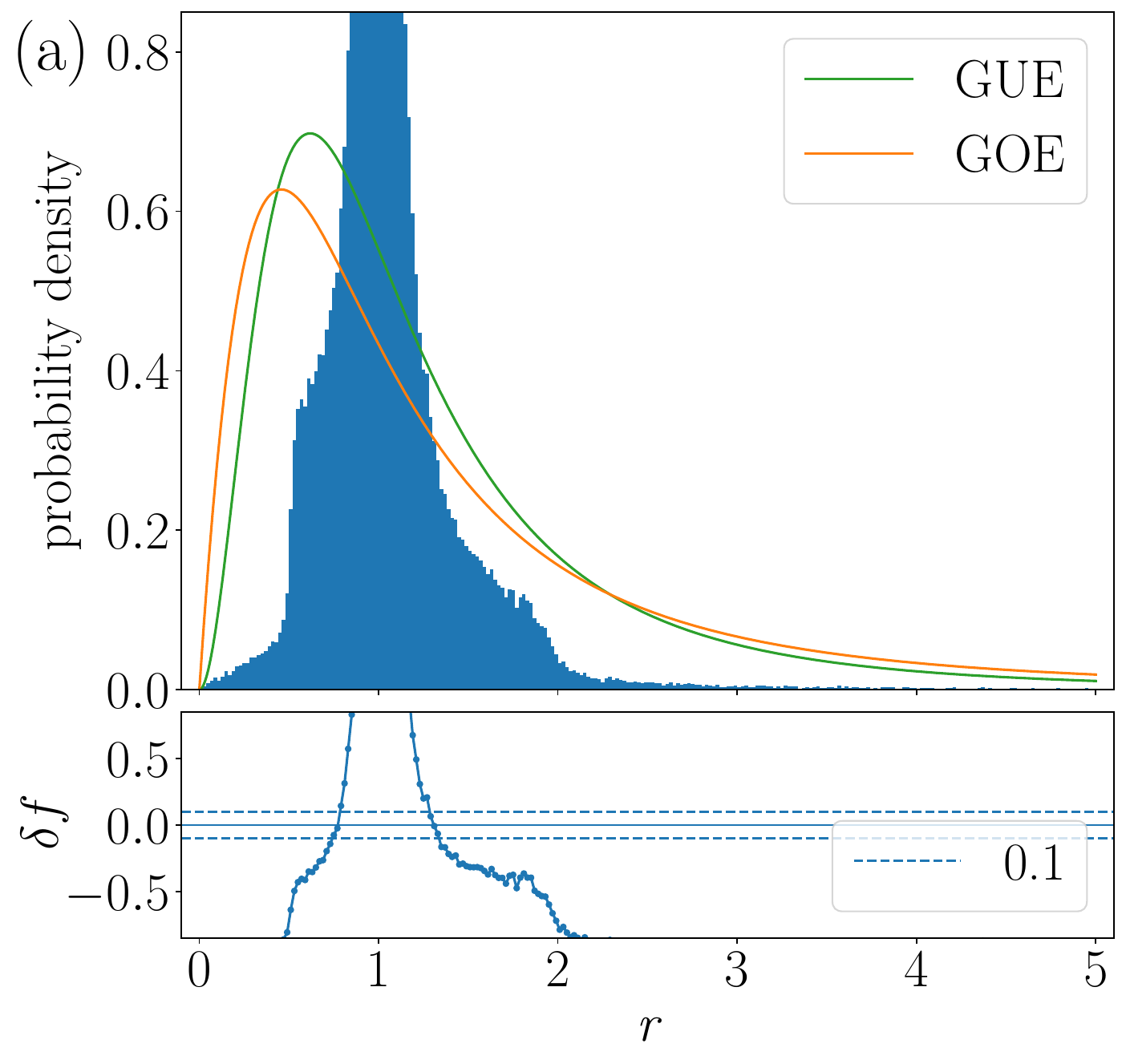}
    \includegraphics[width=0.45\columnwidth]{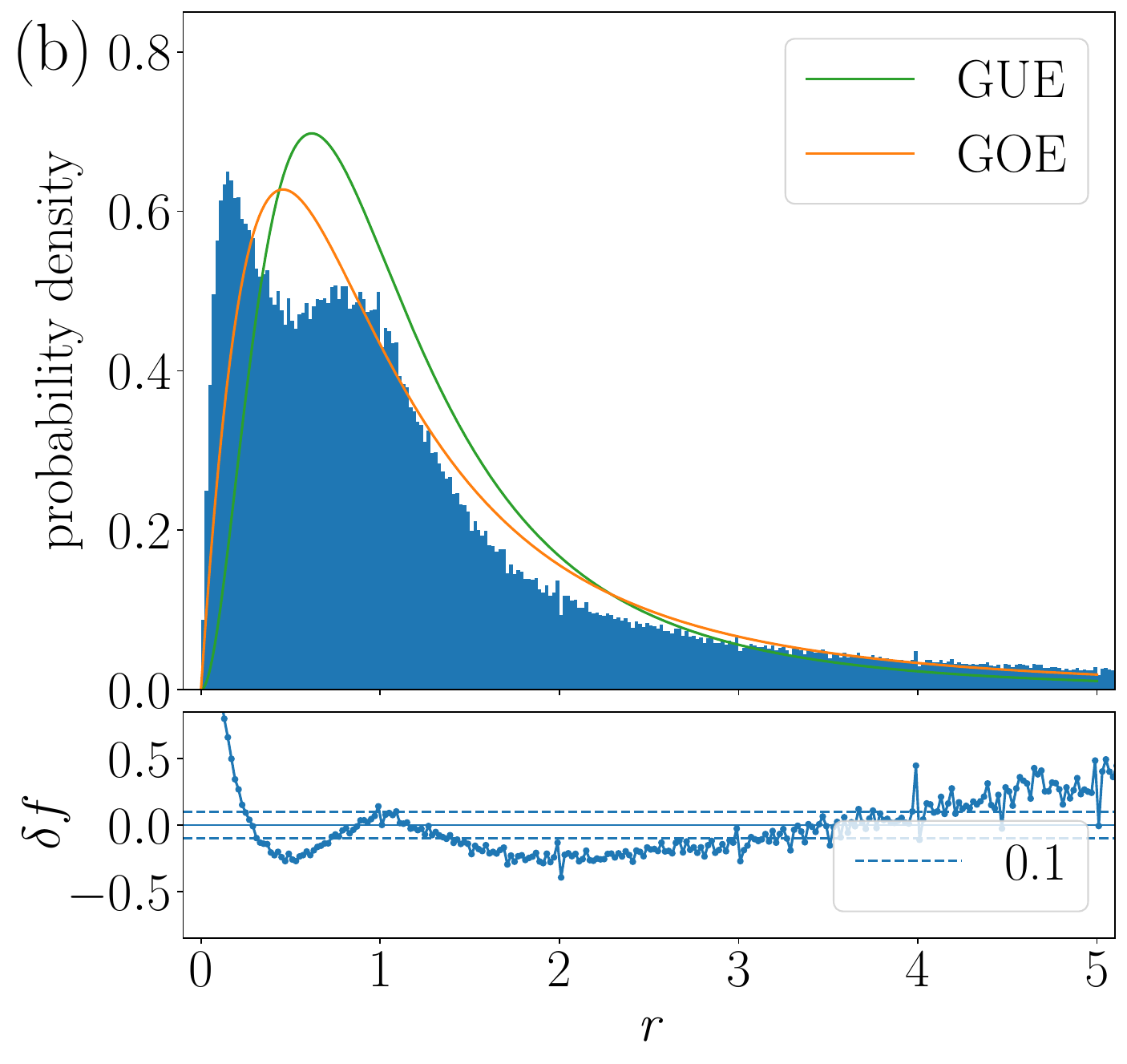}
    \includegraphics[width=0.45\columnwidth]{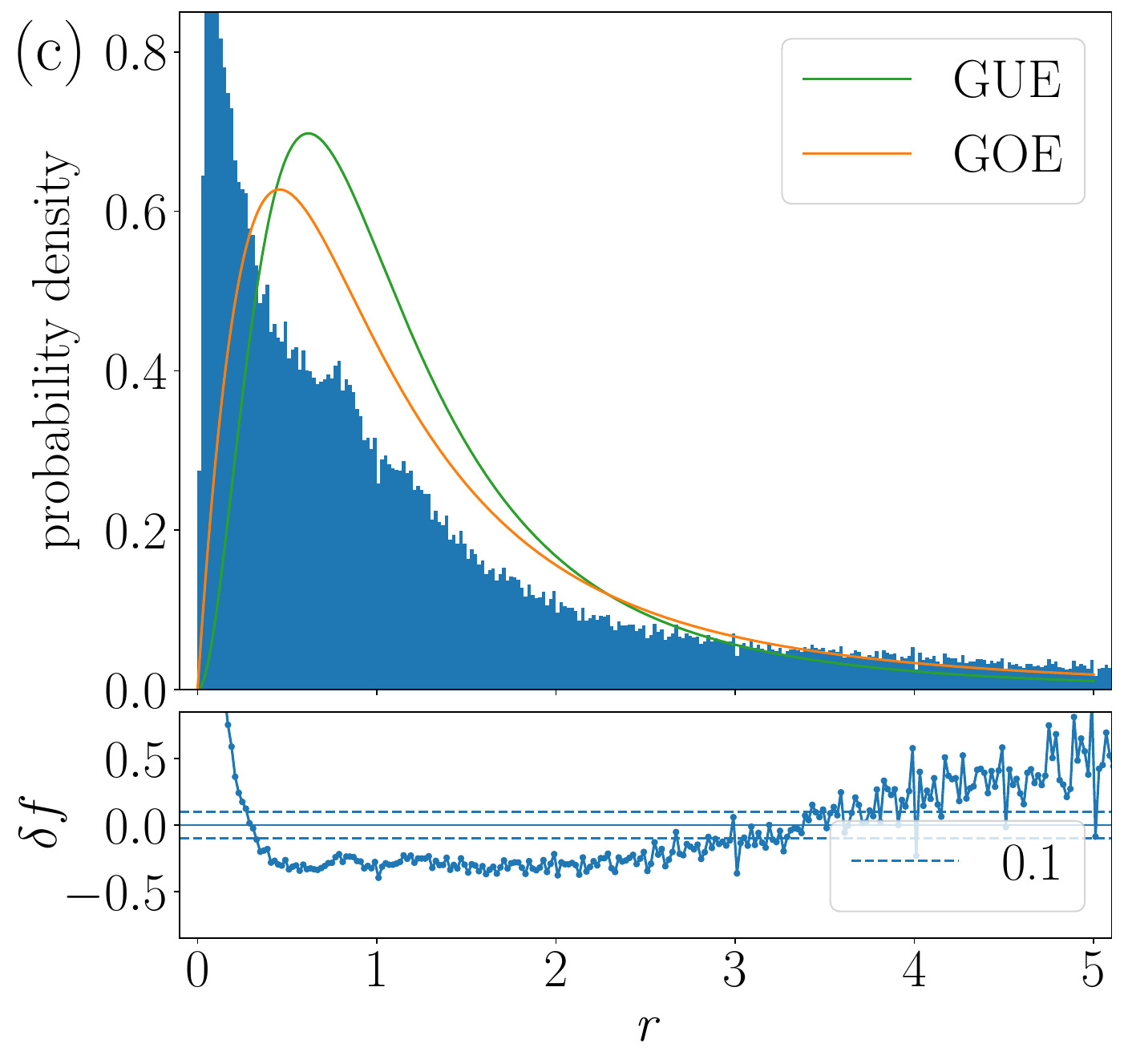}\\
    \includegraphics[width=0.45\columnwidth]{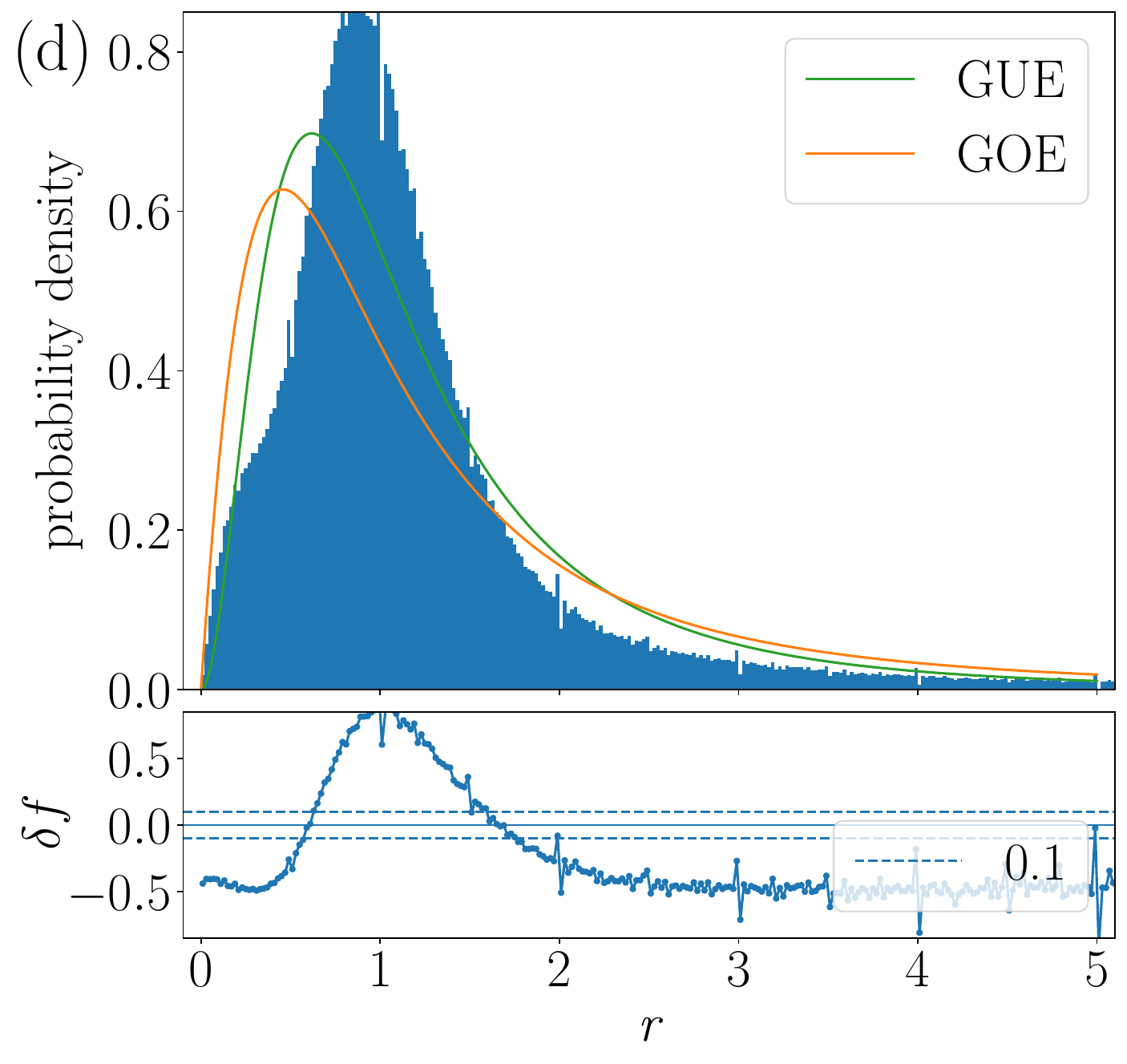}
    \includegraphics[width=0.45\columnwidth]{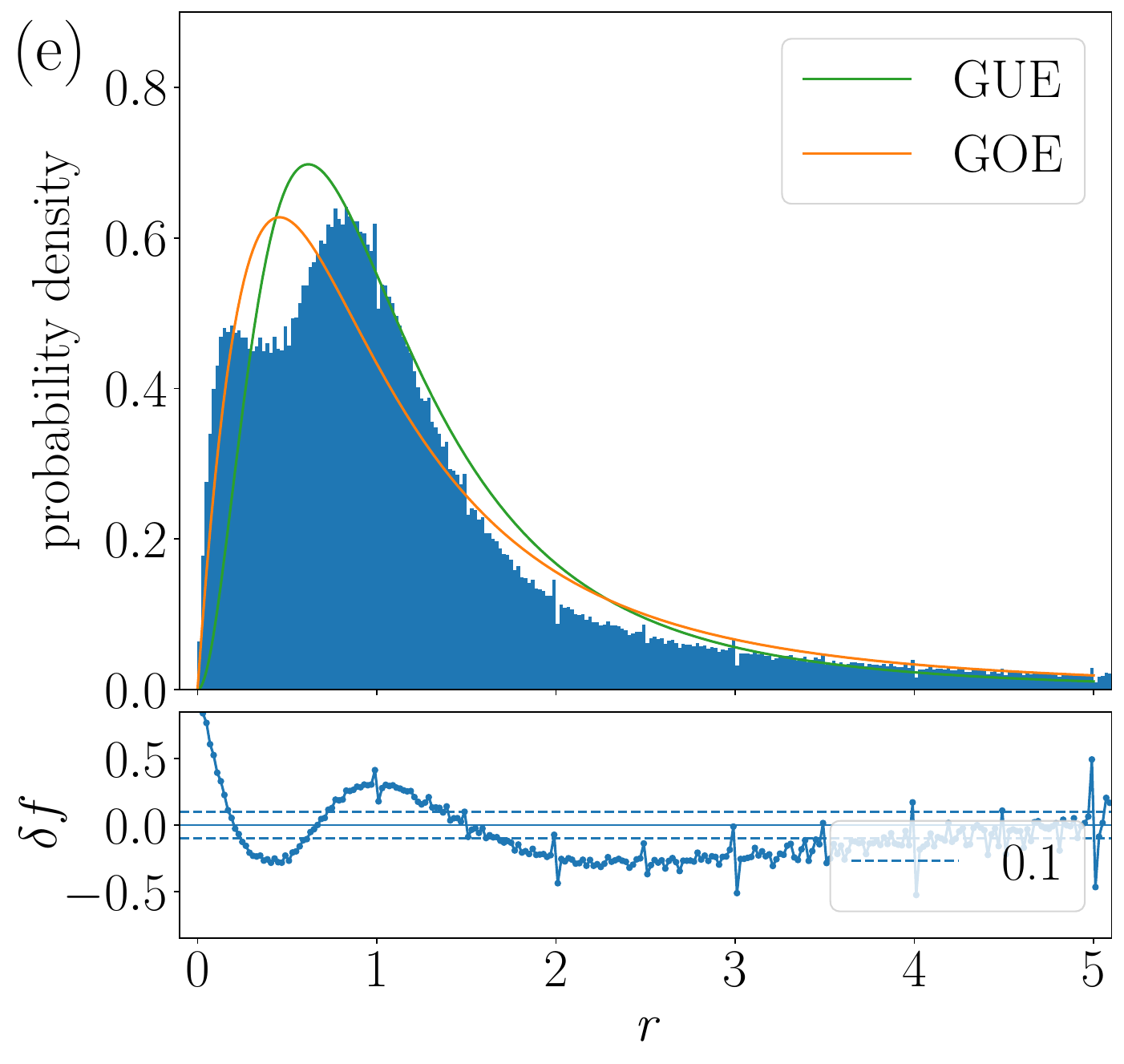}
    \includegraphics[width=0.45\columnwidth]{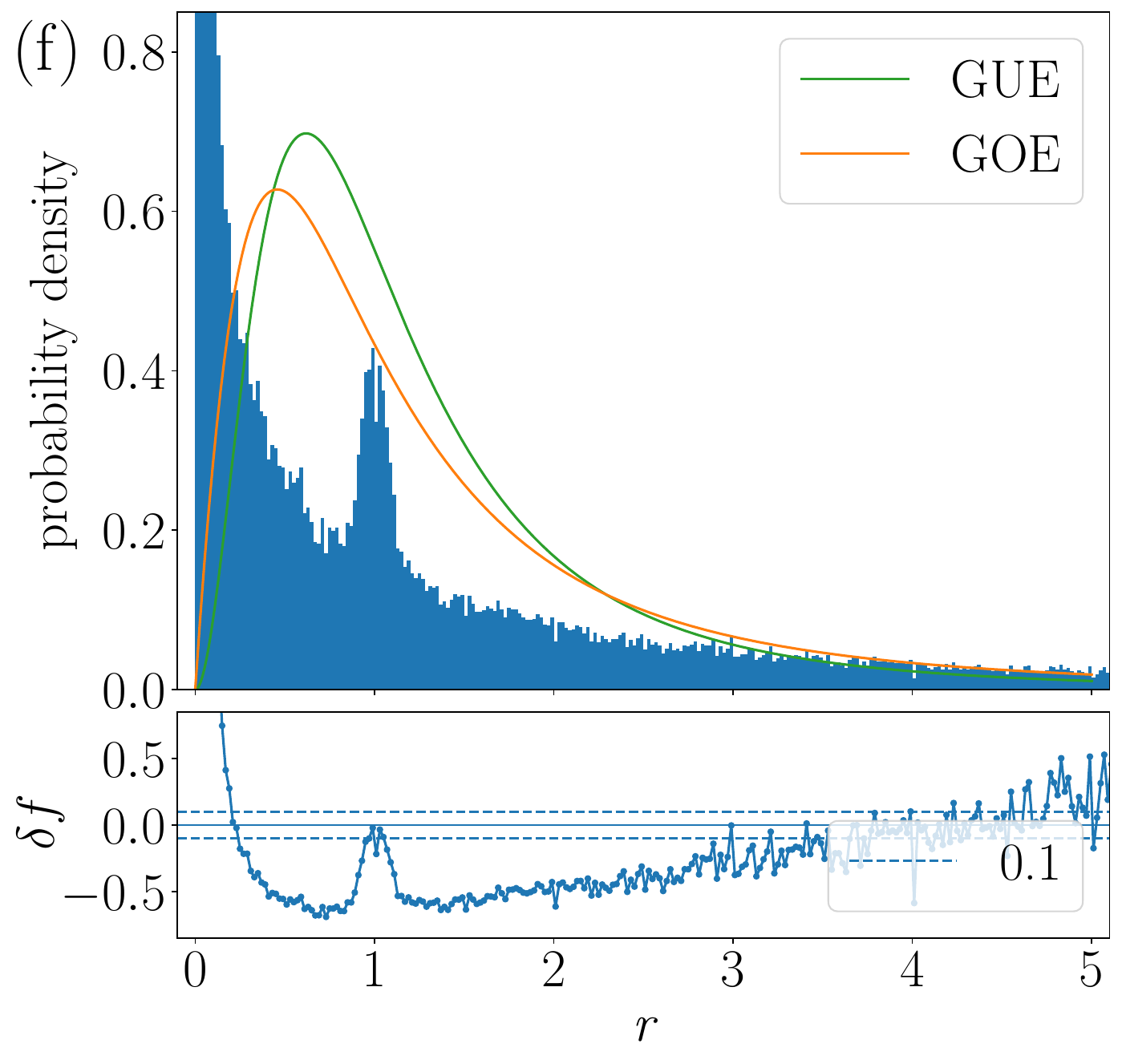}\\
    \includegraphics[width=0.45\columnwidth]{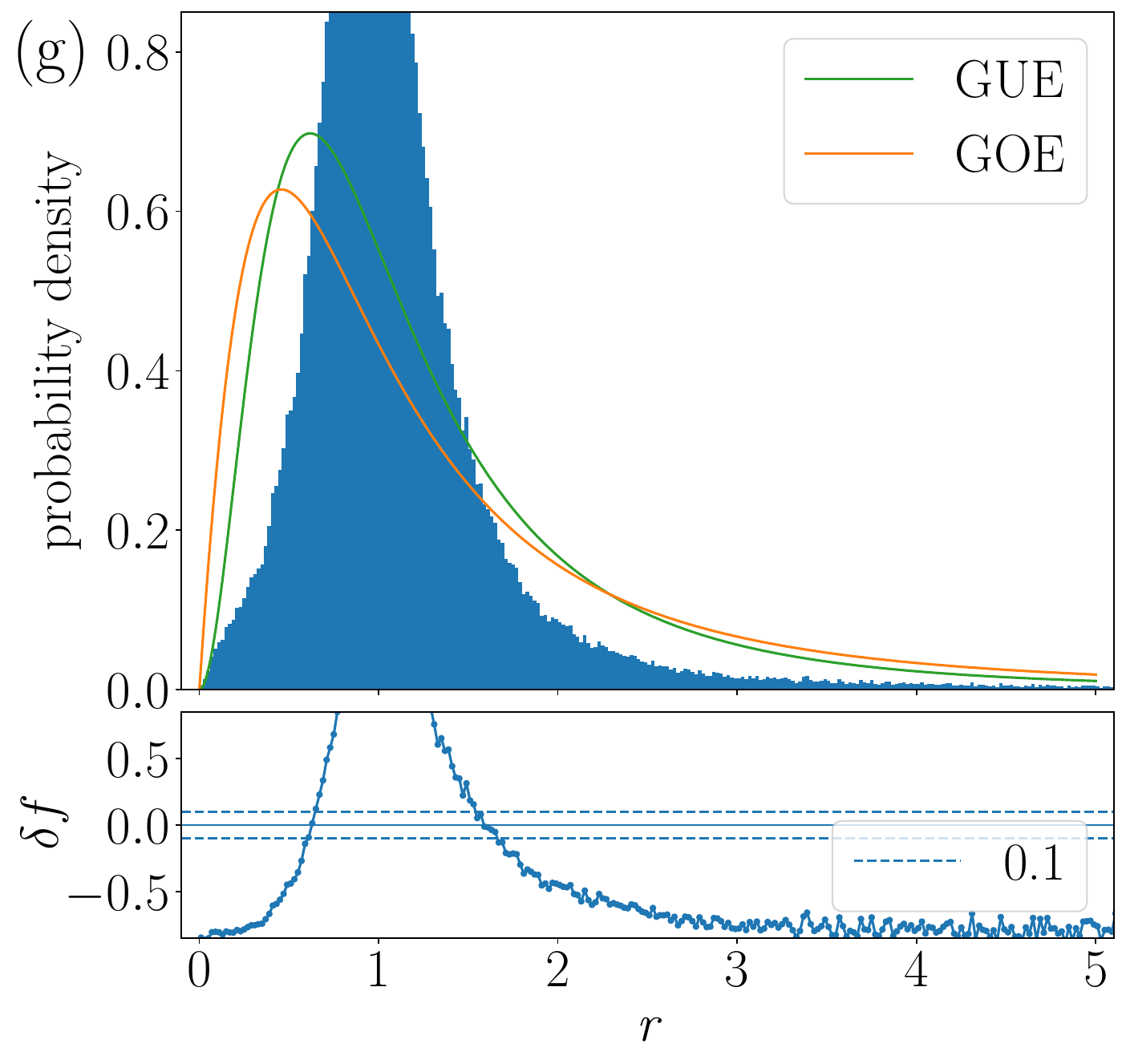}
    \includegraphics[width=0.45\columnwidth]{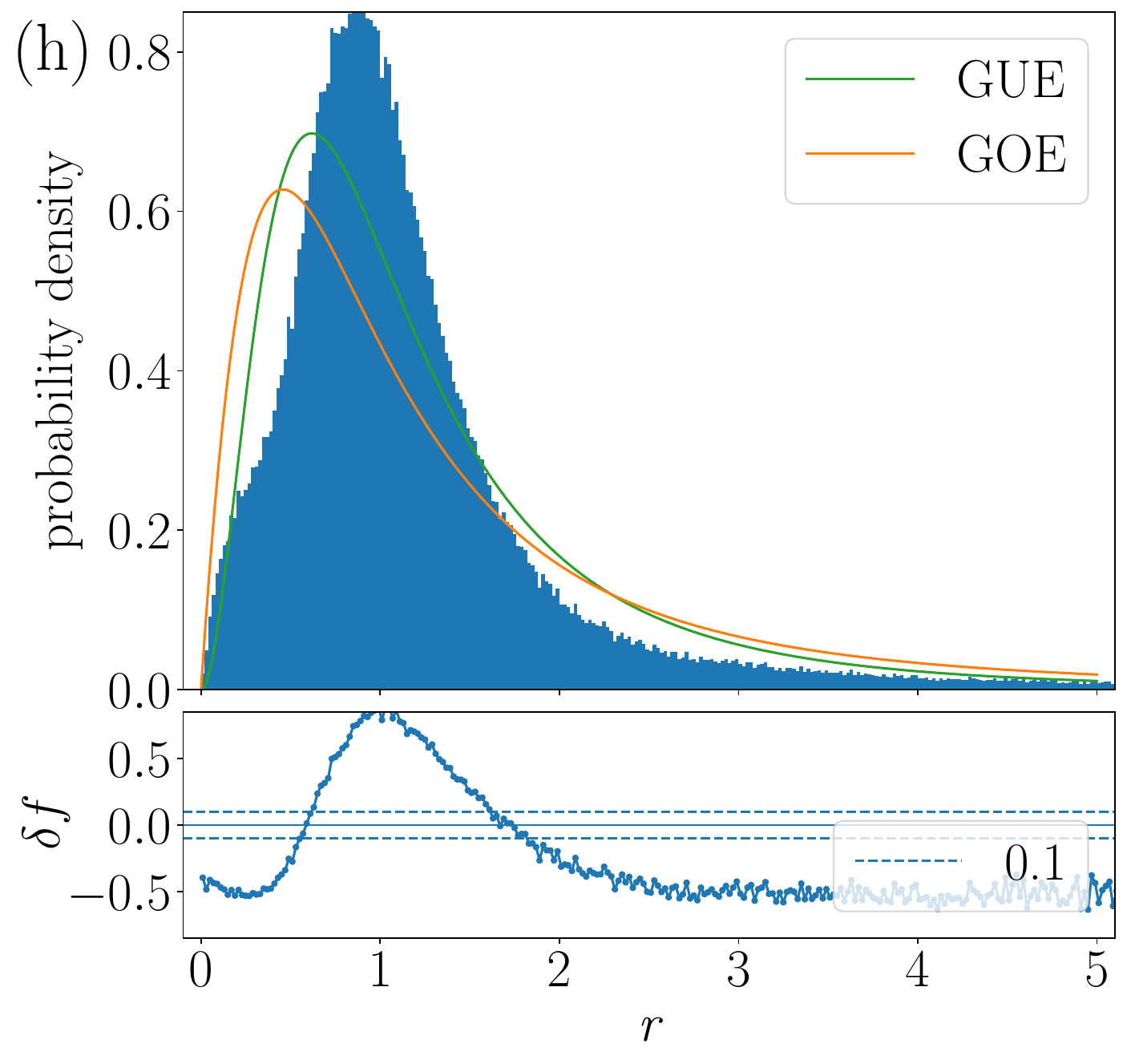}
    \includegraphics[width=0.45\columnwidth]{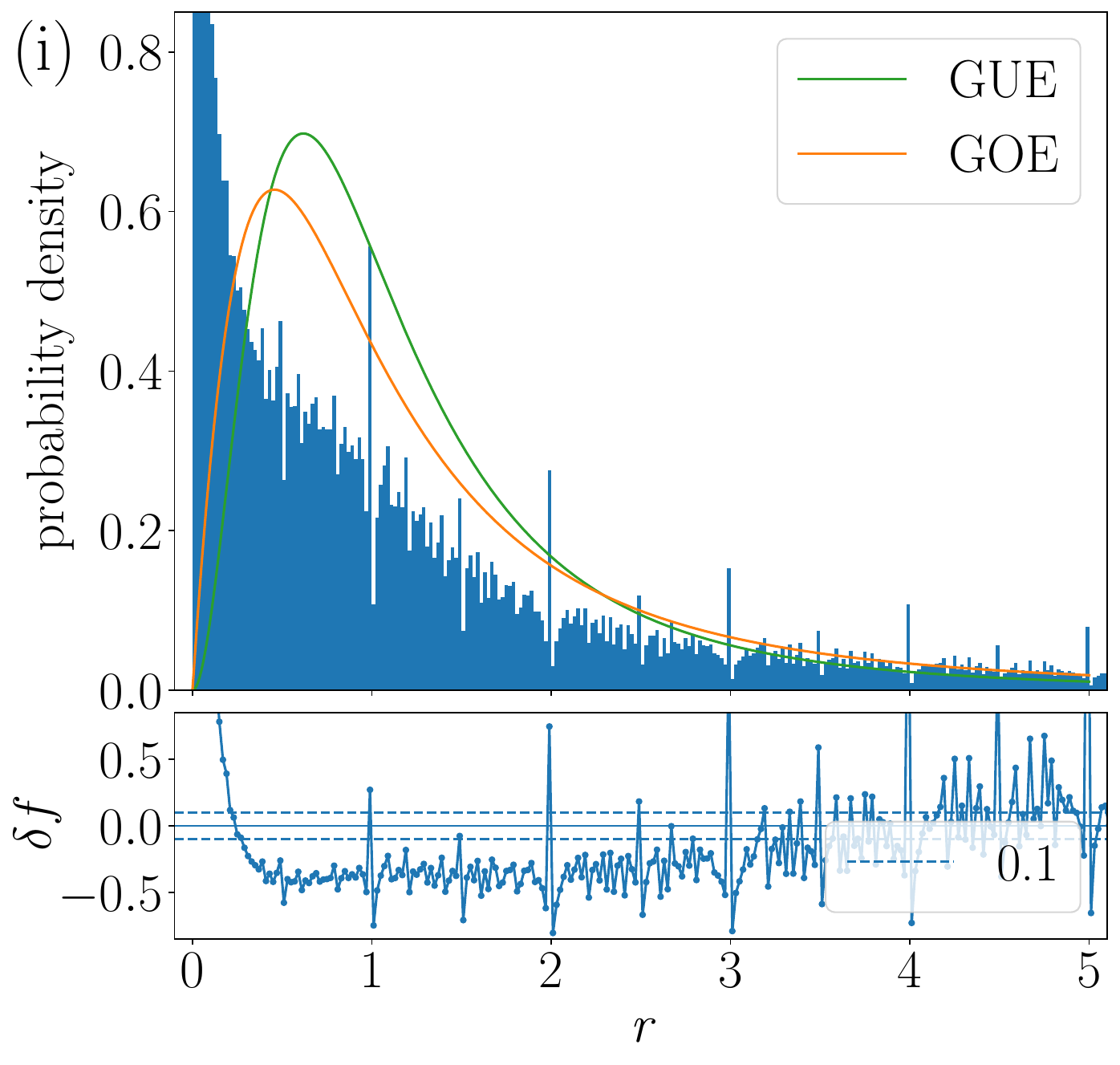}
    \caption{\label{fig:dep_glob}Collective statistics of the spacing ratios of extrema in the global-quench setup for parameter values different from the default values: \textit{(a)}~larger circumference, $L = 100$ ($\Delta x = 1$); \textit{(b)}~larger circumference, $L = 10$ ($\Delta x = 0.1$); \textit{(c)}~smaller circumference, $L = 0.5$; \textit{(d)}~larger mass, ${m = 50}$; \textit{(e)}~larger mass, ${m = 25}$; \textit{(f)}~smaller mass, ${m = 2}$; \textit{(g)}~larger slab width, ${\tau = 0.1}$; \textit{(h)}~larger slab width, ${\tau = 0.05}$; \textit{(i)}~smaller slab width, ${\tau = 0.001}$. The default values of the parameters are: ${N = 250}$, ${N_t = 5\cdot10^6}$, ${t_f = 500}$, ${\Delta x = 0.01}$, ${L = 1}$, ${m = 10}$, and ${\tau = 0.01}$. The bottom panels show the relative deviation between the data and the GOE PDF.}
\end{figure*}

In the global quench setup, the model parameters are the mass $m$, the circumference $L$, and the slab width $\tau_0$. Across the range considered in \figref{fig:dep_glob}, the distributions remain qualitatively different from the GOE PDF, interpolating between small-ratio contribution and a peak near unity. In this case, crossovers out of the intermediate window in parameter space lead correspondingly to one of these two limiting distributions.

\textit{Weighted-modes form factor.---}In addition to the form factor based on extrema statistics (the scattering form factor of~\cite{Bianchi:2024fsi}), one can consider other state-dependent diagnostics to address the ambiguity in identifying the relevant spectrum in a quantum field theory with an operator insertion.

By analogy with the use of statistics of spacetime correlation patterns, the operator spectral weights associated with the field modes can be treated as effective spectral data. This viewpoint is also close in spirit to recent studies~\cite{Das:2022evy, Jeong:2024jjn, Ageev:2024gem} in which the effective spectral data are built from probe-field normal modes on curved-spacetime backgrounds.

In the local-quench example, the operator spectral weights are the coefficients~$\{c_n\}$ arising from the mode decomposition of the field operator. For the finite-volume case, the locally excited state can be written as
\be
    \ket{\Psi(t)} = e^{-iH(t - t_0)}\sum_n c_n\ket{n},
\ee
where $\ket{n}$ are the Hamiltonian eigenstates and
\be
    c_n = \left(\sum_k\frac{e^{-2\eps\om_k}}{2\om_k}\right)^{-\frac{1}{2}}\frac{e^{-\eps\om_n}}{\sqrt{2\om_n}}.
    \label{eq:weights}
\ee
These coefficients allow us to construct a form factor directly from the definition
\be
    g(t) = g_0\sum_{i,\,j\,=\,1}^{N} e^{-i(\lambda_i - \lambda_j)t},
\ee 
by replacing $\{\lambda_i\}$ with $\{c_n\}$ and taking $\tilde{t}$ as the dimensionless Fourier-conjugate variable. We call this version the weighted-modes form factor.

\begin{figure*}
    \includegraphics[width=0.65\columnwidth]{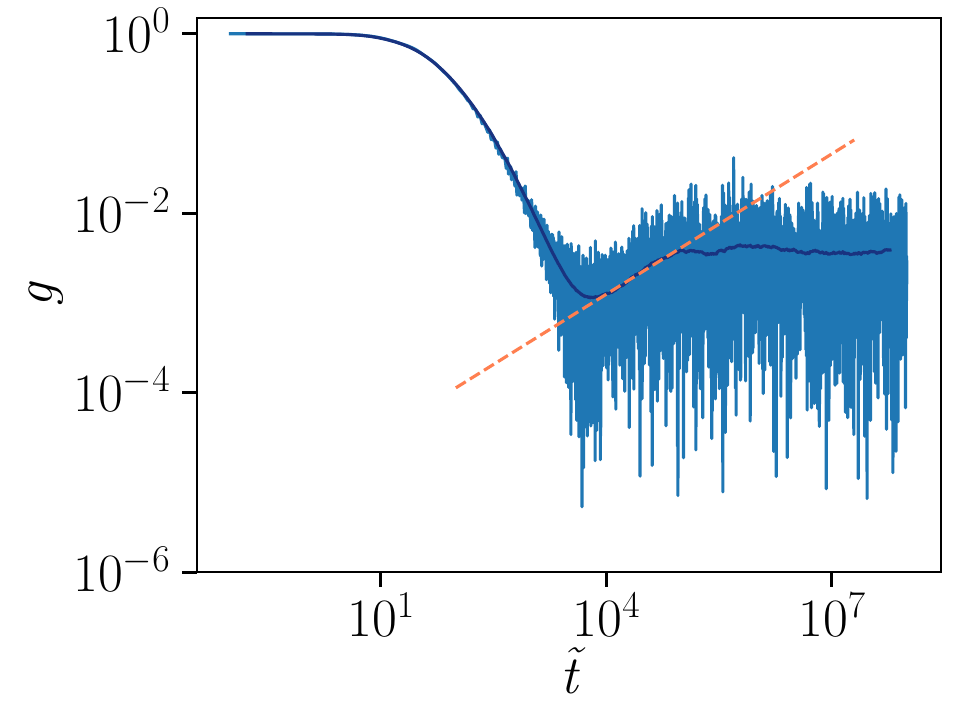}
    \caption{\label{fig:FF_modes}Weighted-modes form factor for the local operator quench. The parameters are: ${L = 1}$, ${m = 10}$, ${\eps = 10^{-3}}$, and ${N = 250}$. The darker line shows the moving average of the data; the dashed orange line shows a linear fit to the averaged curve over the ramp region.}
\end{figure*}

\figref{fig:FF_modes} shows that the weighted-modes form factor exhibits a ramp for suitable parameter choices. Specifically, the choice of $\eps$ determines which cutoff values of $N$ yield a ramp: smaller $\eps$ allows larger $N$. The cutoff value must also be large enough to capture the essentially nonlinear part of the operator spectral weight sequence. Under variations in $m$, the structure is much more stable.

A parameter analogous to the inverse temperature can be included in the definition of the SFF~\cite{Cotler:2016fpe},
\be
    \begin{aligned}
        g(\beta, t) &= g_0\sum_{i,\,j\,=\,1}^{N}  e^{-\beta(\lambda_i + \lambda_j)}e^{-i(\lambda_i - \lambda_j)t},
    \end{aligned}
\ee 
serving as a damping that smooths the curve. The same modification can be made for the weighted-modes form factor. The fit to the moving-averaged data in \figref{fig:FF_modes} does not have exactly unit slope, but there exists a value of $\beta$ for which it does.

\bibliography{main_arXiv}

\end{document}